%% file: main.tex
\documentclass[11pt]{article}

\usepackage[preprint]{acl}

\usepackage{times}
\usepackage{latexsym}
\usepackage{makecell}

\usepackage[T1]{fontenc}

\usepackage[utf8]{inputenc}

\usepackage{microtype}

\usepackage{inconsolata}

\usepackage{graphicx}

\usepackage{booktabs}       
\usepackage{multirow}
\usepackage{multicol}
\usepackage{colortbl}
\usepackage[table]{xcolor}

\usepackage{utfsym}

\usepackage{booktabs}
\usepackage{multirow}
\usepackage{graphicx}
\usepackage{enumitem}
\usepackage{amsmath} 

\title{Rank4Gen: RAG-Preference-Aligned Document Set Selection and Ranking}

\author{
Yongqi Fan\textsuperscript{\rm $\diamondsuit$\thanks{~~Co-first authors.}},
Yuxiang Chu\textsuperscript{\rm $\diamondsuit$\footnotemark[1]},
Zhentao Xia\textsuperscript{\rm $\diamondsuit$},
 Xiaoyang Chen\textsuperscript{\rm $\heartsuit \clubsuit$},
 Jie Liu\textsuperscript{\rm $\spadesuit$}, \\
 \textbf{Haijin Liang}\textsuperscript{\rm $\spadesuit$},
 \textbf{Jin Ma}\textsuperscript{\rm $\spadesuit$},
 \textbf{Ben He}\textsuperscript{\rm $\heartsuit \clubsuit$},
 \textbf{Yingfei Sun}\textsuperscript{\rm $\heartsuit$},
 \textbf{Jie Zhai}\textsuperscript{\rm $\diamondsuit$}\thanks{~~Corresponding author.},
 \textbf{Dezhi Ye}\textsuperscript{\rm $\spadesuit$}\footnotemark[2],
 \textbf{Tong Ruan}\textsuperscript{\rm $\diamondsuit$}\footnotemark[2] \\
\textsuperscript{\rm $\diamondsuit$}East China University of Science and Technology, Shanghai, China \\
\textsuperscript{\rm $\spadesuit$}Tencent, \textsuperscript{\rm $\heartsuit$}University of Chinese Academy of Sciences \\
\textsuperscript{\rm $\clubsuit$}Chinese Information Processing Laboratory, Institute of Software, CAS \\
\texttt{johnnyfans@mail.ecust.edu.cn}, \texttt{dezhiye@tencent.com}, \texttt{ruantong@ecust.edu.cn} \\
}

\begin{document}
\maketitle
\begin{abstract}
In the RAG paradigm, document ranking determines the evidence available to downstream generators. Through controlled analysis, we identify two phenomena underexplored by existing rankers: (i) downstream response quality depends not only on relevance but also on the composition and ordering of selected documents, and (ii) such preferences differ systematically across generators. However, existing rankers are trained purely on query--document relevance, leaving both phenomena unmodeled. To close this gap, we construct \textbf{PRISM}, a bilingual preference-aligned dataset built through a four-stage pipeline that compresses the combinatorial subset-and-ordering space by roughly four orders of magnitude and produces response-quality preference supervision conditioned on seven downstream generators. On a 13k-query subset of PRISM, we train \textbf{Rank4Gen}, a generator-aware ranker that performs joint document set selection and ordering. Experiments on five challenging RAG benchmarks show that Rank4Gen improves downstream QA quality on most evaluated generators, with per-generator F1 gains of up to $+2.08$ over the strongest set-selection baseline. Code is available at \url{https://github.com/JOHNNY-fans/Rank4Gen}.
\end{abstract}

\section{Introduction}

\begin{figure}[t]
    \centering
    \includegraphics[width=\linewidth]{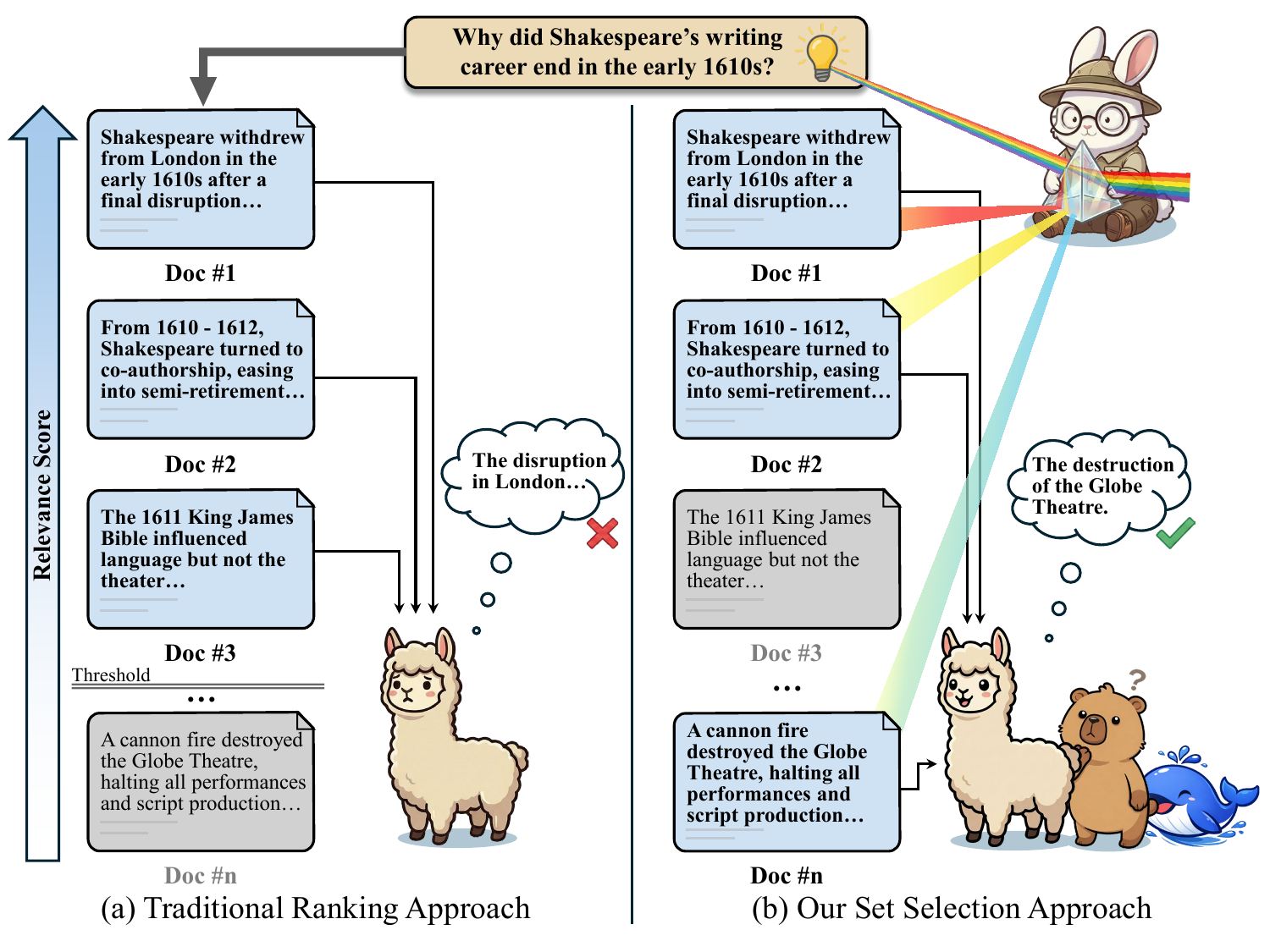}
    \caption{Comparison between traditional ranking and Rank4Gen in RAG. Traditional relevance-based ranking may truncate crucial evidence, while Rank4Gen performs generator-aware document set selection.}
    \label{fig:intro}
\end{figure}

\begin{figure*}[t]
    \centering
    \includegraphics[width=1\linewidth]{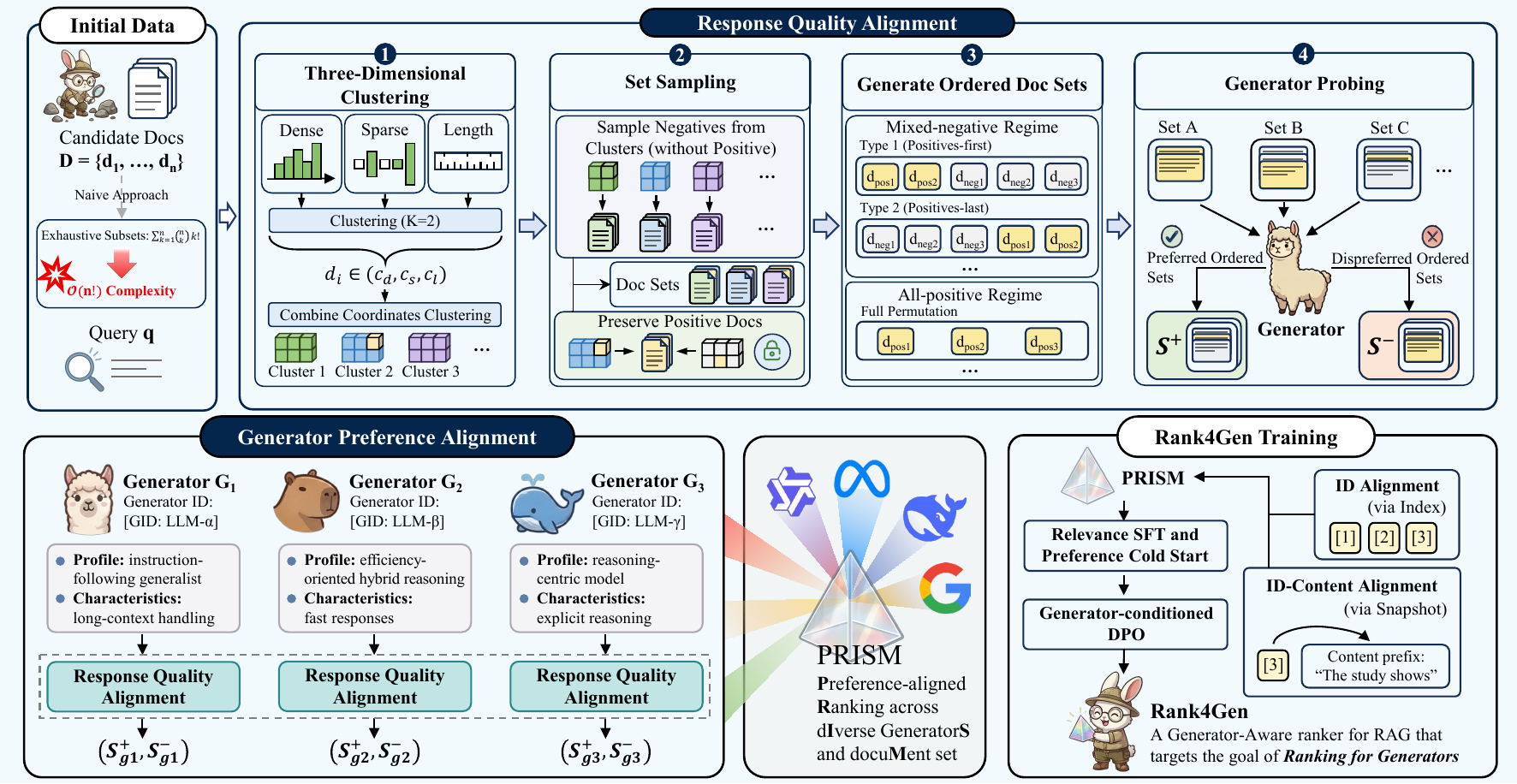}
    \caption{Overview of PRISM and Rank4Gen. PRISM is constructed by (1) Response Quality Alignment, which derives preferred and dispreferred ordered document sets per generator, and (2) Generator Preference Alignment, which aggregates such pairs across diverse generators. Rank4Gen is then trained on PRISM with SFT and generator-conditioned DPO under \texttt{/index} and \texttt{/snapshot} output modes.}
    \label{fig:framework}
\end{figure*}

Retrieval-Augmented Generation (RAG) grounds large language model (LLM) outputs in external knowledge, mitigating hallucinations in knowledge-intensive tasks~\citep{lewis2020retrieval, izacard2021leveraging, huang2025survey}. Since retrieved documents directly determine the evidence available to the generator, ranking quality is critical to the overall effectiveness of RAG.

Most existing RAG systems rely on retrieval and ranking models optimized for query--document relevance, such as dense retrievers, neural rerankers, and more recent LLM-based ranking approaches~\citep{karpukhin2020dense, nogueira2019passage, sun2023chatgpt, fan2025tfrank}. While relevance-based ranking is effective for identifying documents related to the query, relevance alone does not fully reflect generators' preferences for evidence usage during generation. Recent studies show that retrieved documents containing correct answers may still fail to support correct generation, whereas seemingly less relevant documents can sometimes better facilitate reasoning and answer synthesis~\citep{tian2025relevance}. As a result, the mismatch between retrieval relevance and generation utility limits the effectiveness of ranking improvements on downstream response quality.

Moreover, most existing ranking approaches assume a generator-agnostic setting, applying a single ranking strategy regardless of the downstream generator. In practice, different generators vary in how they utilize retrieved context, which can cause document selection and ordering strategies to generalize poorly across generators, leading to unstable performance~\citep{liu2024lost,laban2024summary}. Recent work has explored set-level formulations and LLM-based ranking approaches for RAG to better support evidence aggregation across multiple documents~\citep{pradeep2023rankzephyr,liu2025reasonrank,lee2025shifting}. However, these methods remain largely optimized for relevance-based objectives and do not explicitly align ranking decisions with downstream generation quality or account for generator-specific preferences.

To address these issues, we pursue the goal of \emph{Ranking for Generators}. We first probe how downstream generators consume retrieved evidence and identify two consistent patterns: response quality is shaped by both subset composition and ordering, and the preferred arrangement varies markedly across generators. To make these phenomena learnable, we construct \textbf{PRISM} (Preference-aligned Ranking across dIverse generatorS and docuMent sets), a bilingual preference-aligned dataset built via a four-stage pipeline that operationalizes two key principles: (1) \textbf{From Ranking Relevance to Response Quality}, supervising ranking by downstream response quality rather than query--document relevance; and (2) \textbf{Generator-Specific Preference Modeling}, conditioning supervision on each downstream generator to capture their distinct preferences. Building on PRISM, we train \textbf{Rank4Gen}, a generator-aware ranker that adopts a set selection paradigm (Figure~\ref{fig:intro}), with relevance SFT followed by generator-conditioned DPO~\citep{rafailov2023direct}.

Across five challenging RAG benchmarks, we compare Rank4Gen with relevance-based, LLM-based, and set-selection baselines and conduct ablation and generalization analyses, observing downstream gains on most of the evaluated generators.

Our contributions are summarized as follows:
\begin{itemize}
    \item We conduct an empirical analysis of RAG preference phenomena, showing that downstream response quality depends not only on document relevance but also on subset composition and ordering, and that these preferences differ systematically across generators.
    \item Building on these observations, we construct \textbf{PRISM}, a bilingual preference-aligned dataset that provides per-generator response-quality supervision over both the composition and ordering of candidate documents across seven downstream generators.
    \item We propose \textbf{Rank4Gen}, a generator-aware ranker trained on \textbf{PRISM\_13K}, a 13k-query subset of PRISM, with relevance SFT and generator-conditioned DPO, and show through extensive experiments on five challenging RAG benchmarks that it improves downstream generation quality on most evaluated generators.
\end{itemize}

\section{Related Work}
\subsection{Ranking Methods and Paradigms}
Traditional information retrieval ranking methods typically follow pointwise, pairwise, or listwise paradigms. Pointwise methods independently estimate the relevance of each document to a given query~\citep{nogueira2019passage}. Pairwise methods compare pairs of documents to infer their relative ordering~\citep{qin2024large}, while listwise methods model the entire candidate document list jointly~\citep{tang2024listwise}. With the rapid development of LLMs, recent work has explored leveraging their strong capabilities for ranking. Some approaches directly employ LLMs via prompting to score candidate documents~\citep{sun2023chatgpt}. Other methods focus on distillation, transferring the ranking ability of LLMs into smaller models, such as RankZephyr~\citep{pradeep2023rankzephyr}. More recent studies further introduce reinforcement learning, optimizing ranking policies using rewards derived from ranking metrics or preference signals~\citep{sun2025grouprank, sun2025dynamicrag}.

\subsection{LLM for Ranking in RAG}
Early RAG systems construct generation contexts via fixed Top-K retrieval or adaptive depth selection~\citep{lewis2020retrieval, jeong2024adaptive}. Subsequent work incorporates LLMs for reranking, either to improve ranking effectiveness~\citep{yu2024rankrag, zhang2025rearank} or to enhance reasoning efficiency and reduce inference cost~\citep{reddy2024first, liu2025leveraging}. Other studies reformulate ranking as document subset selection to control better context composition~\citep{meng2025ranking, lee2025shifting, zhang2025distilling}. Despite these advances, most methods optimize relevance-based objectives rather than downstream generation quality, and approaches such as RankRAG~\citep{yu2024rankrag} unify ranking and generation in a single LLM rather than learning a standalone ranker for multiple downstream generators. Recent work has begun exploring joint optimization of rankers and generators~\citep{shi2025direct}.

\begin{figure*}[t]
    \centering
    \includegraphics[width=0.8\linewidth]{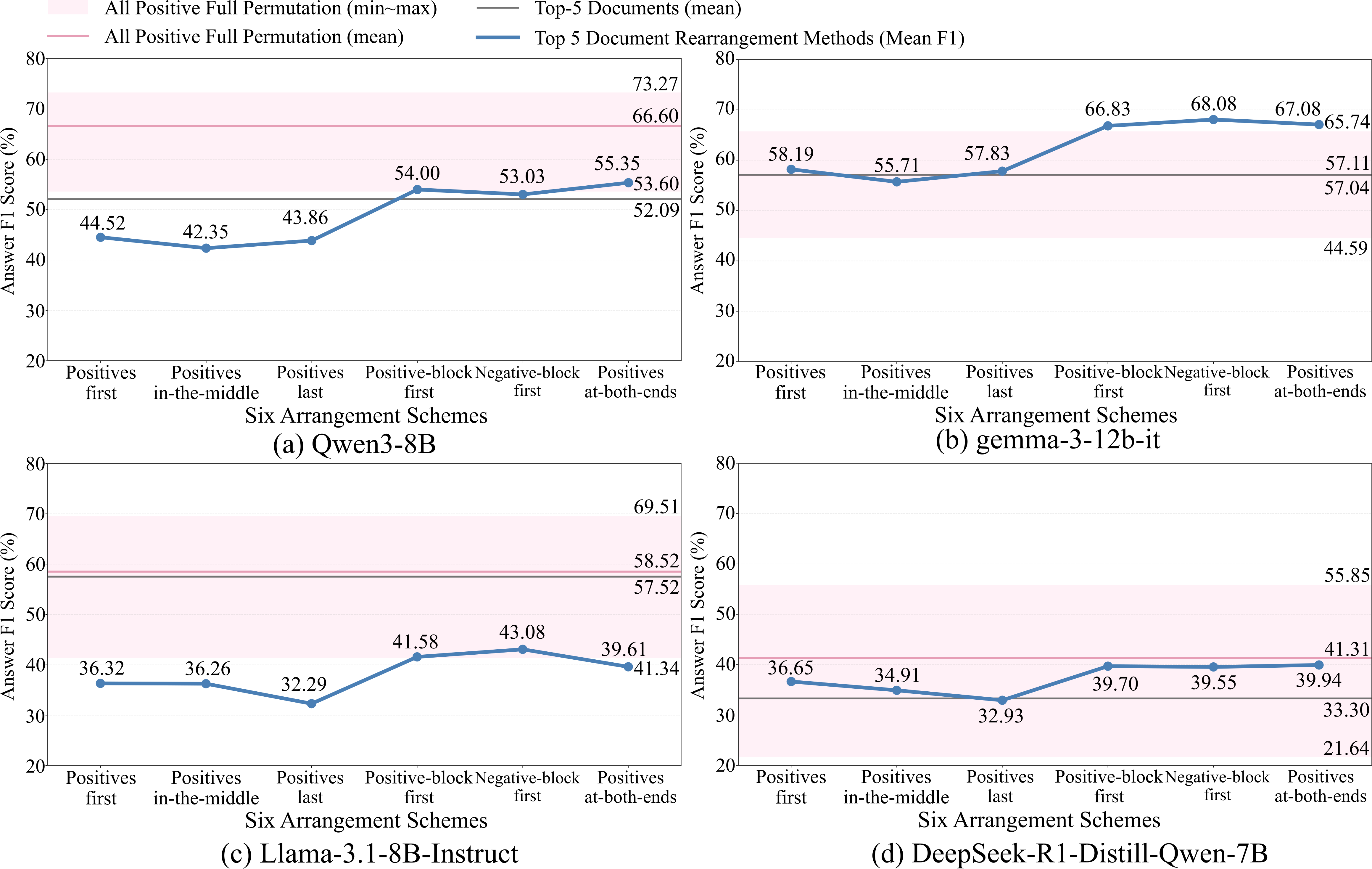}
    \caption{Impact of document subset composition and ordering on RAG performance across generators.}
    \label{fig:phenomenon}
\end{figure*}

\section{PRISM Dataset}
\subsection{Task Definition}
In the RAG setting, we consider ranking as a generator-aware document set selection problem. Given a query $q$, some candidate documents $\mathcal{D} = \{d_1, \ldots, d_N\}$, and a downstream generator $G$, the goal of the ranker is to select a document subset and produce an ordering over it. The resulting ordered subset $\mathcal{S} = (d_{i_1}, d_{i_2}, \ldots, d_{i_k})$, where $\mathcal{S} \subseteq \mathcal{D}$, is directly provided to the generator as contextual evidence for RAG. Under this formulation, our ranker adopts a set selection paradigm that outputs an ordered set of documents tailored to the downstream generator.

\subsection{Preference Phenomena in RAG}
\label{sec:phenomenon}
To motivate the dataset design, we first conduct a controlled study that varies document subset composition and ordering across downstream generators.

For each query, we use bge-m3~\citep{chen2024bge} to retrieve the top-5 candidate documents and construct multiple ordered document subsets under two complementary settings. The \emph{all-positive} setting enumerates all permutations of the positive set, isolating preferences over the ordering of relevant evidence. The \emph{mixed-negative} setting retains all positives and adds sampled negatives, where the joint space of orderings is too large to enumerate; we instead probe it with six predefined positional schemes (Figure~\ref{fig:schemes} in Appendix~\ref{app:prism_details}), each targeting a recurring LLM long-context regime---primacy bias~\citep{liu2024lost}, recency bias, lost-in-the-middle, and endpoint anchoring. Every ordered subset is provided to a downstream generator under a fixed QA prompt and evaluated against the gold answer.

Figure~\ref{fig:phenomenon} summarizes the results across multiple downstream generators. Different document subsets and orderings lead to substantial performance variations even when all documents are individually relevant, indicating that RAG performance depends not only on relevance but also on evidence composition and ordering. Moreover, preferences vary across generators: subsets that perform well for one generator may perform poorly for another. These observations motivate the generator-aware formulation above and inform PRISM construction below.

\subsection{Curation Methodology}
\label{sec:prism_construction}
To support Rank4Gen, we construct \textbf{PRISM}, a preference-aligned dataset that captures both response-quality preferences over ordered document subsets and per-generator preferences across seven downstream generators.

As shown at the top of Figure~\ref{fig:framework}, PRISM is built through a four-stage pipeline: (i)~corpus collection; (ii)~three-dimensional fusion clustering of candidate documents; (iii)~cluster-guided subset sampling and phenomenon-aware positional arrangement; and (iv)~response-quality judging and generator preference alignment. The core challenge is that surfacing a generator's preferences over ordered subsets is expensive: for $n$ candidates the joint subset-selection and ordering space has size $\sum_{k=1}^{n}\binom{n}{k}k!$, and scoring each ordered subset further requires one downstream generator inference plus one judge call, so the dominant cost is the per-subset LLM-evaluation budget. Stages~(ii)--(iii) therefore compress this input space into a small controlled pool along two structural axes—\emph{evidence composition} and \emph{evidence ordering}—keeping the per-(query, generator) judge calls bounded. Stage~(iv) then turns these ordered subsets into the bilingual per-generator preference pairs that supervise Rank4Gen.

\paragraph{Stage 1: Data Collection.}
We aggregate publicly available RAG and multi-document QA datasets: HotpotQA~\cite{yang-etal-2018-hotpotqa}, 2WikiMultiHopQA~\cite{ho-etal-2020-constructing}, MUSIQUE~\cite{trivedi-etal-2022-musique}, MS MARCO~\cite{nguyen2016ms}, and CRUD-RAG~\cite{lyu2025crud}. Each instance is normalized into a query \(q\), a source-provided gold answer \(a_q\), a supporting document set \(\mathcal{P}_q\), and a candidate document pool \(\mathcal{D}_q\). The resulting bilingual source pool contains 141k queries.

\begin{figure*}[t]
    \centering
    \includegraphics[width=0.95\linewidth]{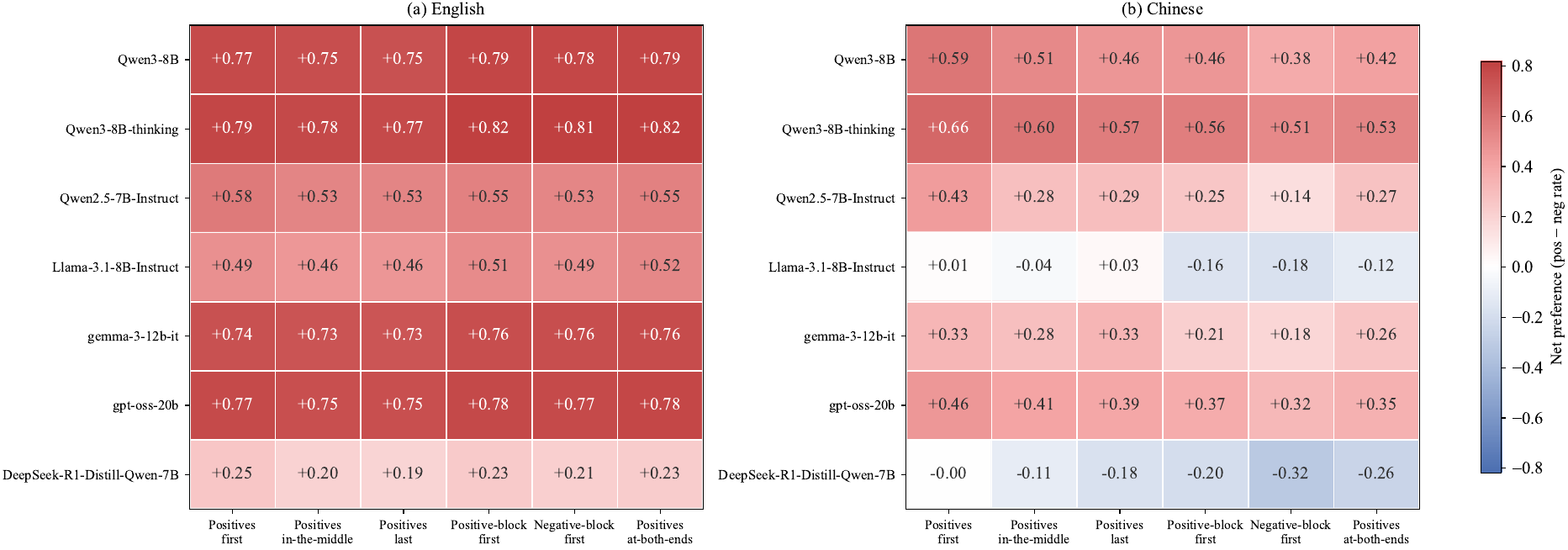}
    \caption{Per-template net preference (preferred $-$ dispreferred rate) of each generator over the six Stage~\hyperref[sec:subset_sampling]{3}(b) templates, on EN (left) and ZH (right). Red: preferred; blue: avoided.}
    \label{fig:fingerprint}
\end{figure*}

\paragraph{Stage 2: Three-Dimensional Fusion Clustering.}
\label{sec:3dcluster}

To avoid repeatedly sampling near-duplicate distractors from the candidate pool \(\mathcal{D}_q\), we assign each document three binary labels along complementary axes: a dense-semantic label, a sparse-lexical label, and a length label.

For the two content-based axes, we apply the same clustering recipe: K-Means with \(k=2\) over L2-normalized representations. Specifically, we use bge-m3 embeddings to obtain a dense-semantic label \(c_d\) and TF-IDF unigram/bigram vectors to obtain a sparse-lexical label \(c_s\). For the length axis, we set \(c_\ell = \mathbf{1}[\log(1{+}|d|) > m_q]\), where \(|d|\) is the word count of \(d\) and \(m_q\) is the per-query median, separating short snippets from long-form passages. Each candidate document \(d \in \mathcal{D}_q\) is thus assigned a cluster code
\begin{equation}
\phi(d) = (c_d,\, c_s,\, c_\ell) \in \{0,1\}^3.
\end{equation}

Documents with the same \(\phi(d)\) are placed in the same cluster. This exact-match fusion produces at most \(2^3=8\) clusters per query and encourages sampled negatives from different clusters to vary in topic, surface form, or length; the empirical per-query cluster-count distribution is reported in Appendix~\ref{app:prism_details}.

\paragraph{Stage 3: Subset Sampling and Positional Arrangement.}
\label{sec:subset_sampling}
Given the diversity clusters, Stage~3 produces a controlled set of ordered subsets that vary along the two structural axes targeted by PRISM: \emph{evidence composition} (which documents are included) and \emph{evidence ordering} (in what order).

\noindent\textbf{(a) Cluster-Guided Combination Sampling.}
For each query we compose a small number of unordered document bags that Stage~3(b) will later arrange into ordered subsets. Every bag contains all positives in \(\mathcal{P}_q\) and diverse negatives drawn from the Stage~2 clusters, with at most one negative per cluster so that distractors avoid near-duplicates. We aim for roughly twice as many negatives as positives but cap the bag size at \(M_{\max}=8\), so the number of negatives per bag is
\begin{equation}
r_q = \min\!\bigl(2 n_q^{+},\ M_{\max}-n_q^{+},\ |\mathcal{C}_q^{-}|\bigr),
\end{equation}
where \(n_q^{+}=|\mathcal{P}_q|\) and \(\mathcal{C}_q^{-}\) is the set of clusters containing at least one negative. The three terms encode, respectively, the target negative-to-positive ratio, the per-bag document budget, and the available diversity. We repeat this sampling \(K=3\) times with a fixed random seed and remove duplicate bags, yielding up to three unordered subsets per query.

\noindent\textbf{(b) Phenomenon-Aware Positional Sampling.}
Rather than enumerating the $m_q!$ orderings per subset, we directly apply the six phenomenon-targeted positional templates introduced in Section~\ref{sec:phenomenon}; per-template feasibility constraints on $(n^{+},n^{-})$ (Appendix~\ref{app:prism_details}) reduce some bag-template pairs, yielding at most $K\cdot 6=18$ ordered subsets per query.

\paragraph{Stage 4: Listwise Judging and Pair Extraction.}
\label{sec:stage4_judging}

\noindent\textbf{Probing inputs.} Mirroring Section~\ref{sec:phenomenon}, we instantiate two complementary regimes per query:
\begin{itemize}[leftmargin=*,itemsep=2pt,topsep=2pt]
    \item \emph{All-positive regime}: enumerate all $n_q^{+}!$ permutations of $\mathcal{P}_q$ alone (typically at most $6$, as most queries have $n_q^{+}\leq 3$), isolating preferences over the ordering of relevant evidence.
    \item \emph{Mixed-negative regime}: apply the Stage~\hyperref[sec:subset_sampling]{3}(b) templates to the Stage~\hyperref[sec:subset_sampling]{3}(a) subsets, exposing the joint preference over composition and ordering.
\end{itemize}
We write $\mathcal{Q}_{q}$ for the resulting pool of candidate ordered subsets per query.

\noindent\textbf{Judge labeling.} For each $\mathcal{S}\in\mathcal{Q}_q$, generator $G$ produces an answer; an LLM-as-a-judge then receives the gold answer $a_q$ together with all candidate responses for the query--generator pair $(q,G)$, and returns a listwise ranking with a binary correctness flag per response, scoring both reasoning and final answer; full judge configuration is provided in Appendix~\ref{app:judge_config}.

\noindent\textbf{Pair extraction.} For each $\mathcal{S}\in\mathcal{Q}_q$ we keep its judge label only if the predicted answer text is consistent with the gold answer $a_q$, yielding a preferred pool $\mathcal{S}^{+}_{q,G}$ and a dispreferred pool $\mathcal{S}^{-}_{q,G}$. For each $(q, G)$ we then form a single preference pair $(S^{+}_{q,G}, S^{-}_{q,G})$ from the two pools. Aggregating across queries and generators yields the PRISM preference set
\begin{equation}
\mathcal{T} \;=\; \bigl\{\,(S^{+}_{q,G},\ S^{-}_{q,G}) \,\bigm|\, q,\ G\,\bigr\}.
\end{equation}

To let a single ranker recognize \emph{which} generator it is serving, we further condition each preference tuple on generator-related information: a unique identifier and a textual description summarizing the generator's profile. Stages~\hyperref[sec:subset_sampling]{3}--\hyperref[sec:stage4_judging]{4} are repeated independently for each of the generators, so the resulting tuples inherit a context $(\text{generator id}, \text{description})$.

\paragraph{Dataset statistics and preference signal.}
Running the full pipeline produces \textbf{PRISM}: 141k bilingual queries probed against the seven downstream generators, with $\sim$990k preference pairs extracted in Stage~\hyperref[sec:stage4_judging]{4}. Aggregating mixed-negative judge outputs over PRISM gives the per-template net preference rates in Figure~\ref{fig:fingerprint}: the dominant signal is per-generator \emph{magnitude}, with a $\sim\!0.5$ gap in net preference rate between Qwen3-8B-thinking (top) and DeepSeek-R1-Distill-Qwen-7B (bottom) on both languages, directly justifying generator-conditioned supervision; template-level effects are secondary but visible across languages.

\subsection{Rank4Gen Training}
\label{sec:training}
Rank4Gen is trained on PRISM in two stages: relevance SFT with a preference cold start, followed by generator-conditioned DPO. To strengthen ID--content alignment in the ranker's outputs, both stages share two complementary output modes: \texttt{/index} emits only document IDs, while \texttt{/snapshot} additionally appends the first 100 characters of each document after its ID. Detailed data construction for both stages is provided in Appendix~\ref{app:prism_details}.

\paragraph{Relevance SFT with preference cold start.}
The first stage performs SFT to teach Rank4Gen to identify positive documents from candidates, across two slices: (i) a \texttt{default} slice using each query's positive documents in their original order under a generic LLM description; and (ii) a per-generator slice using the judge's top-ranked all-positive permutation under that generator's profile as gold.

\paragraph{Generator-conditioned DPO.}
The second stage applies DPO on the PRISM preference set $\mathcal{T}$, which spans both the all-positive and mixed-negative regimes (vs. SFT's all-positive-only slice). For each $(S^{+}_{q,G}, S^{-}_{q,G})\in\mathcal{T}$, the ranker is trained to prefer $S^{+}_{q,G}$ over $S^{-}_{q,G}$ under generator $G$'s id and description.

\input{tables/Data_Collection}

\section{Experiment Setup}
\subsection{Datasets and Metrics}
\textbf{Training Dataset.}

We train Rank4Gen on \textbf{PRISM\_13K}, a $12{,}994$-query subset of PRISM sampled under computational constraints. Detailed dataset statistics are reported in Table~\ref{tab:data_overview}.

\textbf{Evaluation Datasets.}
We evaluate Rank4Gen on two groups of RAG benchmarks. The first group focuses on complex reasoning, including multi-hop and temporal question answering, and consists of BrowseComp-Plus~\citep{chen2025browsecomp}, KG-MHQA~\citep{wang2025kg}, and ChronoQA~\citep{chen2025question}. The second group targets factuality evaluation and includes SimpleQA~\citep{wei2024measuring} and ChineseSimpleQA~\citep{he-etal-2025-chinese}, for which we additionally augment candidate documents to enable RAG-based inference.

\textbf{Metrics.}
Following prior work, we evaluate downstream response quality using Exact Match (EM) and F1 scores, computed between the generated responses and the ground-truth answers~\citep{yu2024rankrag}; details in Appendix~\ref{app:metric}.

\input{tables/OOD}

\input{tables/Ablation_table}

\subsection{Baseline}
We evaluate Rank4Gen against a broad set of baselines across seven downstream generators, including LLM-based ranking methods under Pointwise, Listwise, and Set Selection paradigms (see Appendix~\ref{app:baselines} for details). We also include distillation-based ranking methods, including RankZephyr~\citep{pradeep2023rankzephyr} and SETR~\citep{lee2025shifting}, using Gemini-2.5 Flash Lite\footnote{\url{https://docs.cloud.google.com/vertex-ai/generative-ai/docs/models/gemini/2-5-flash-lite}} as the teacher model, and the dense embedding retriever bge-m3~\citep{chen2024bge} as a first-stage retrieval baseline. For fair comparison, all LLM-based methods are trained on the same PRISM\_13K training set.

\subsection{Implementation Details}
We use Qwen3-8B as the backbone. Baselines use two epochs of SFT, while Rank4Gen uses one epoch of SFT followed by one epoch of DPO. All models are trained using ms-swift~\citep{zhao2025swift}, with a learning rate of $1 \times 10^{-5}$ and a warm-up ratio of $0.05$, on 8 NVIDIA H20 GPUs. At inference, the Rank4Gen ranker uses temperature $0$. All downstream generators evaluated in this work follow the same decoding schedule, starting at $T=0$ and only escalating to $0.7$ then $1.0$ when the answer-format parser fails.

\begingroup
\vspace*{-6pt}
\section{Experiment Analysis}
\vspace{-3pt}
\endgroup

\subsection{Main Results}
Table~\ref{tab:main_results} reports the main results on five RAG benchmarks across four representative downstream generators, with full results provided in Table~\ref{tab:full_results}. Overall, Rank4Gen improves downstream response quality on most datasets and generators.

Pointwise and listwise ranking methods achieve competitive performance in certain settings, but their performance is less stable across benchmarks and downstream generators, likely because their reliance on top-$k$ selection from a global ranking limits adaptation to query- and generator-specific subset preferences.

Distillation-based methods show mixed performance across datasets and generators. Their effectiveness appears sensitive to the capability and preference biases of the teacher model, which may not generalize consistently across different RAG benchmarks or generator configurations.

Compared to these baselines, the full Rank4Gen model, trained with relevance SFT and preference DPO and leveraging both \texttt{/index} and \texttt{/snapshot} reasoning modes, achieves improvements on most downstream generators. Relative to the strongest baseline SetSelection-Vanilla, Rank4Gen achieves per-generator F1 gains of $+1.07$ to $+2.08$ across the four representative generators and an overall macro-F1 margin over the other paradigms. These results highlight the benefits of jointly optimizing ranking with respect to response quality and generator-specific preferences.

\input{tables/Generalised_Explanation}

\subsection{Training-Stage and Reasoning-Mode Ablation}

Table~\ref{tab:ablation_rank4gen} presents ablation results for key components of Rank4Gen. By comparing variants trained with different configurations, we analyze the effects of relevance-based initialization, preference optimization, and reasoning modes.

The backbone model handles simple factual queries zero-shot but its document selection does not generalize across generators. Relevance SFT provides a strong foundation for relevance-aware selection, yet performance still varies noticeably across downstream generators under generator-agnostic objectives. Introducing DPO improves overall performance and leads to more stable behavior across generators, highlighting the benefit of answer-quality-based preference optimization.

When combined with the \texttt{/index} mode, the \texttt{/snapshot} reasoning mode provides additional gains in some settings, though its effect is not consistently observed across all datasets.

\subsection{Generalization to Unseen Generators}
We evaluate the generalization ability of Rank4Gen on unseen generators that are not included during PRISM construction. Specifically, we consider an additional open-source LLM and a large-scale proprietary LLM, with details provided in Table~\ref{tab:generators}; despite the name overlap, the unseen DeepSeek-V3.2 is from a different family than the seen DeepSeek-R1-Distill-Qwen-7B, which is distilled from Qwen.

As shown in Table~\ref{tab:generalization}, Rank4Gen lifts these unseen generators from near-zero accuracy without RAG (e.g., Ministral $2.26$ F1 on BrowseComp+) to competitive performance under the default mode, where no generator information is specified, indicating that gains stem from the learned evidence-selection prior rather than intrinsic generator ability. When generator identifiers and descriptions are provided, performance further improves, indicating that Rank4Gen can effectively leverage generator information to better adapt document selection and ranking to new generators.

\section{Conclusion}
In this work, we start from an empirical study: in RAG, downstream response quality is shaped by which documents are selected and how they are ordered, and the preferred arrangement differs markedly across generators. To turn these phenomena into supervision, we build \textbf{PRISM}, a bilingual preference-aligned dataset whose four-stage pipeline yields response-quality preference pairs conditioned on seven downstream generators. Trained on a 13k-query subset, \textbf{Rank4Gen} unifies set selection and ordering in one generator-aware ranker, achieving downstream gains on most evaluated generators across five RAG benchmarks and transferring to unseen generators in default mode.

\section*{Limitations}
This work has a few limitations, primarily related to data scale, preference optimization behavior, and data requirements.

Due to computational constraints, Rank4Gen is trained on a sampled subset of PRISM, namely \textbf{PRISM\_13K}. While our experiments show consistent improvements under this setting, training on the full PRISM dataset may further enhance model capacity and robustness. We plan to release the full PRISM dataset to support future research.

From the evaluation results after DPO training, we observe that downstream F1 scores tend to improve, while Exact Match (EM) may decrease for some generators, as shown in Table~\ref{tab:main_results}. This suggests a trade-off introduced by preference optimization with mixed positive and negative document subsets, which can encourage the ranker to select larger or more diverse document sets. As a result, generators may produce longer reasoning processes and more diverse responses, improving partial correctness without always achieving exact matching.

In addition, PRISM construction relies on datasets with annotated positive and negative documents. Applying our data construction pipeline to fully unlabeled corpora would require additional large language model–based annotation modules, which may introduce additional annotation cost and limit scalability in such settings.

Finally, we represent each generator using a fixed identifier and a textual description generated through a RAG-based pipeline. While manual correction mitigates potential hallucination artifacts, this static representation may not fully capture the generator’s context-dependent evidence preferences. Future work could explore more dynamic generator representations, for example, by allowing the ranker to interact with the generator before ranking to infer its characteristics from contextual signals and adapt document selection accordingly.

\section*{Ethical Considerations}
This work uses only publicly available datasets and models that can be accessed through official sources, and all such resources are properly cited. The construction of PRISM does not involve human annotation, and all preference signals are derived automatically through model-based evaluation. During the course of this project, AI assistants were used solely to support auxiliary tasks such as coding assistance and text polishing. These tools were used in a responsible manner and did not replace human judgment in research design, experimentation, or result interpretation.

\bibliography{custom}

\clearpage

\appendix

\renewcommand{\thetable}{A\arabic{table}}
\renewcommand{\thefigure}{A\arabic{figure}}
\setcounter{figure}{0}
\setcounter{table}{0}

\section{Additional Dataset and Experiment Details}


\subsection{Generator Configuration}

\begin{figure*}[t]
    \centering
    \includegraphics[width=\linewidth]{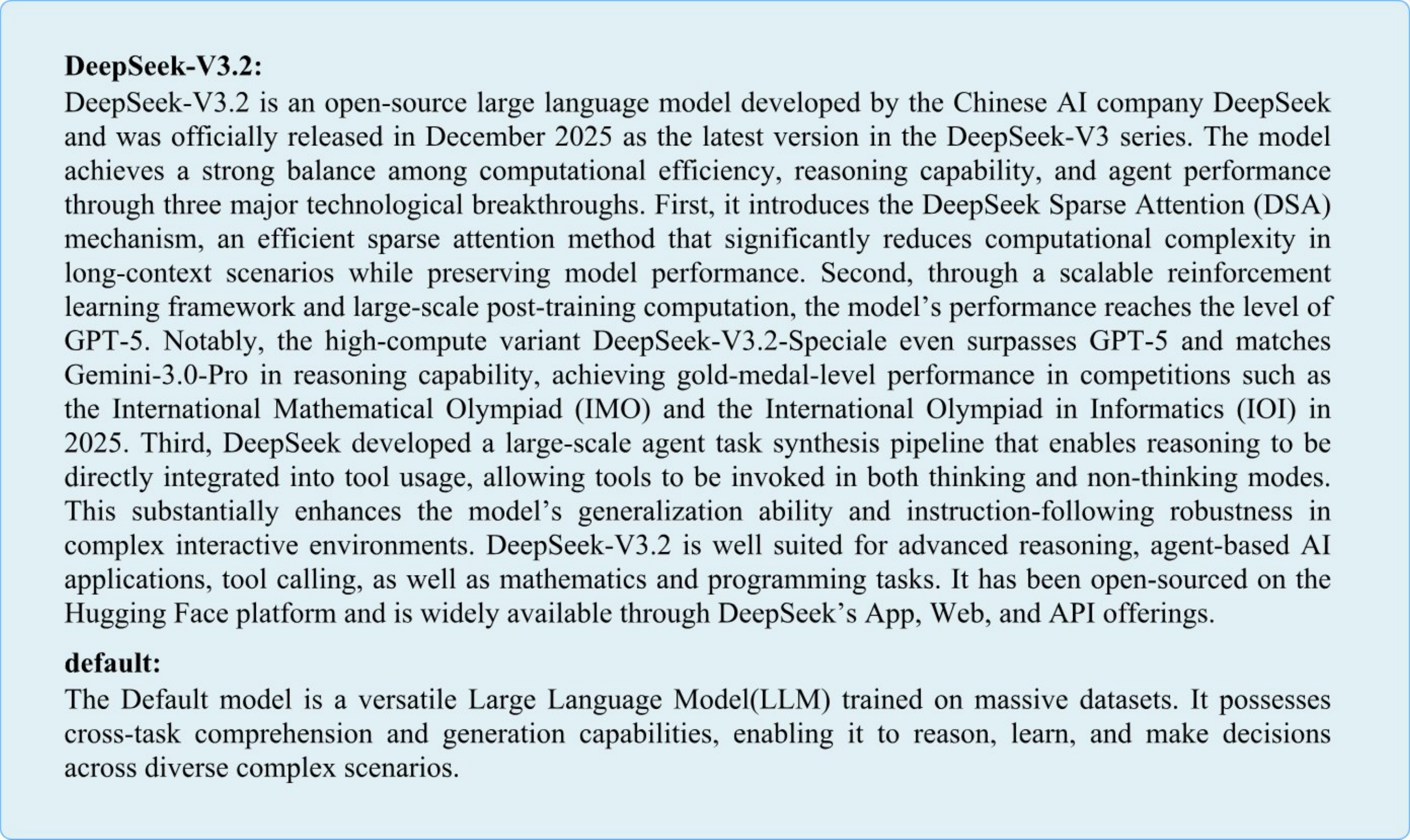}
    \caption{Examples of descriptions for each generator (en).}
    \label{fig:d1}
\end{figure*}

\begin{figure*}[t]
    \centering
    \includegraphics[width=\linewidth]{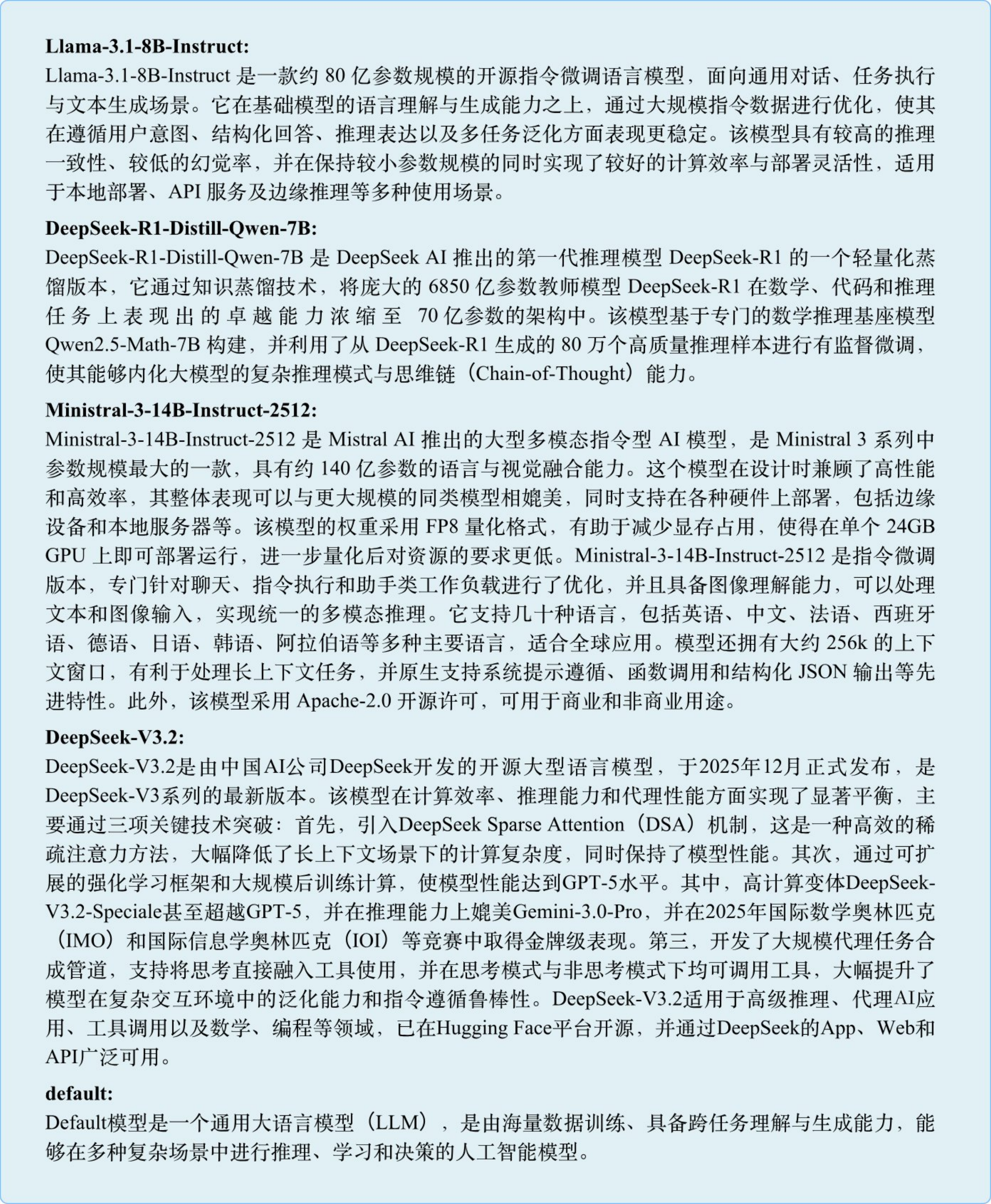}
    \caption{Examples of descriptions for each generator (zh).}
    \label{fig:d2}
\end{figure*}

Table~\ref{tab:generators} lists the generators considered in this work, including those within PRISM, which are used during PRISM construction and Rank4Gen training, and those without PRISM, which are excluded and only used for evaluating generalization to unseen generators. Figures~\ref{fig:d1} and~\ref{fig:d2} present example generator descriptions in English and Chinese, respectively.

\input{tables/generators}

\input{tables/appendex_table}

\subsection{Baseline Description}
\label{app:baselines}
We provide detailed training setups on PRISM\_13K for three basic ranking paradigms: pointwise, listwise, and set selection.

\paragraph{Pointwise Ranking.}
The pointwise paradigm treats document ranking as a binary relevance classification problem.
Each candidate document is independently scored with respect to the query using a binary relevance label (e.g., yes/no),
where documents annotated as positive in PRISM\_13K are treated as relevant.
The ranker is trained to predict the relevance of each document individually, and documents are ranked by their predicted scores.

\paragraph{Listwise Ranking.}
The listwise paradigm takes the full set of candidate documents as input and directly outputs an ordering over all candidates.
During training, the ranker is optimized to promote positive documents to higher positions in the output ranking, while documents are otherwise ordered according to their original retrieval order when no preference is specified.

\begin{figure}[!t]
    \centering
    \includegraphics[width=\linewidth]{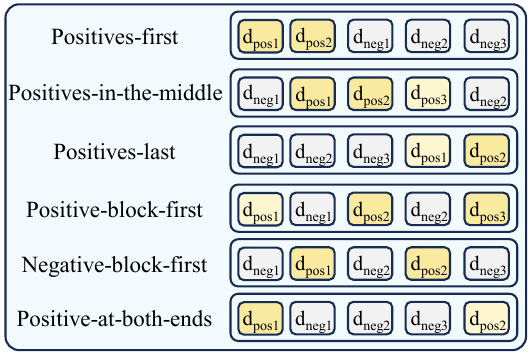}
    \caption{The six predefined positional templates used in Section~\ref{sec:phenomenon} and Stage~3(b).}
    \label{fig:schemes}
\end{figure}

\begin{figure}[!t]
    \centering
    \includegraphics[width=\linewidth]{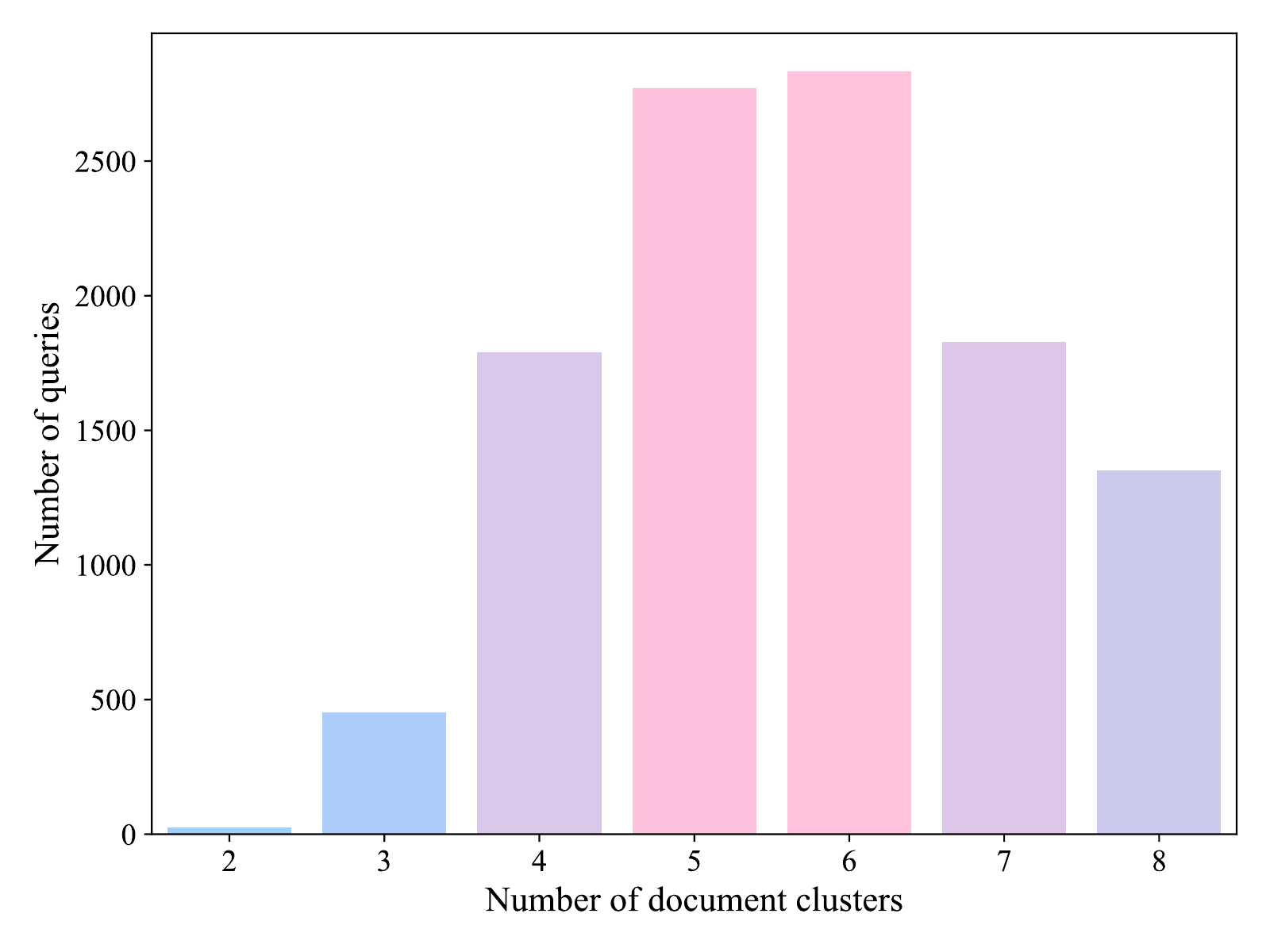}
    \caption{Distribution of the number of clusters derived from the fused three-dimensional coordinate representation.}
    \label{fig:cluster}
\end{figure}

\begin{figure*}[t]
    \centering
    \includegraphics[width=\linewidth]{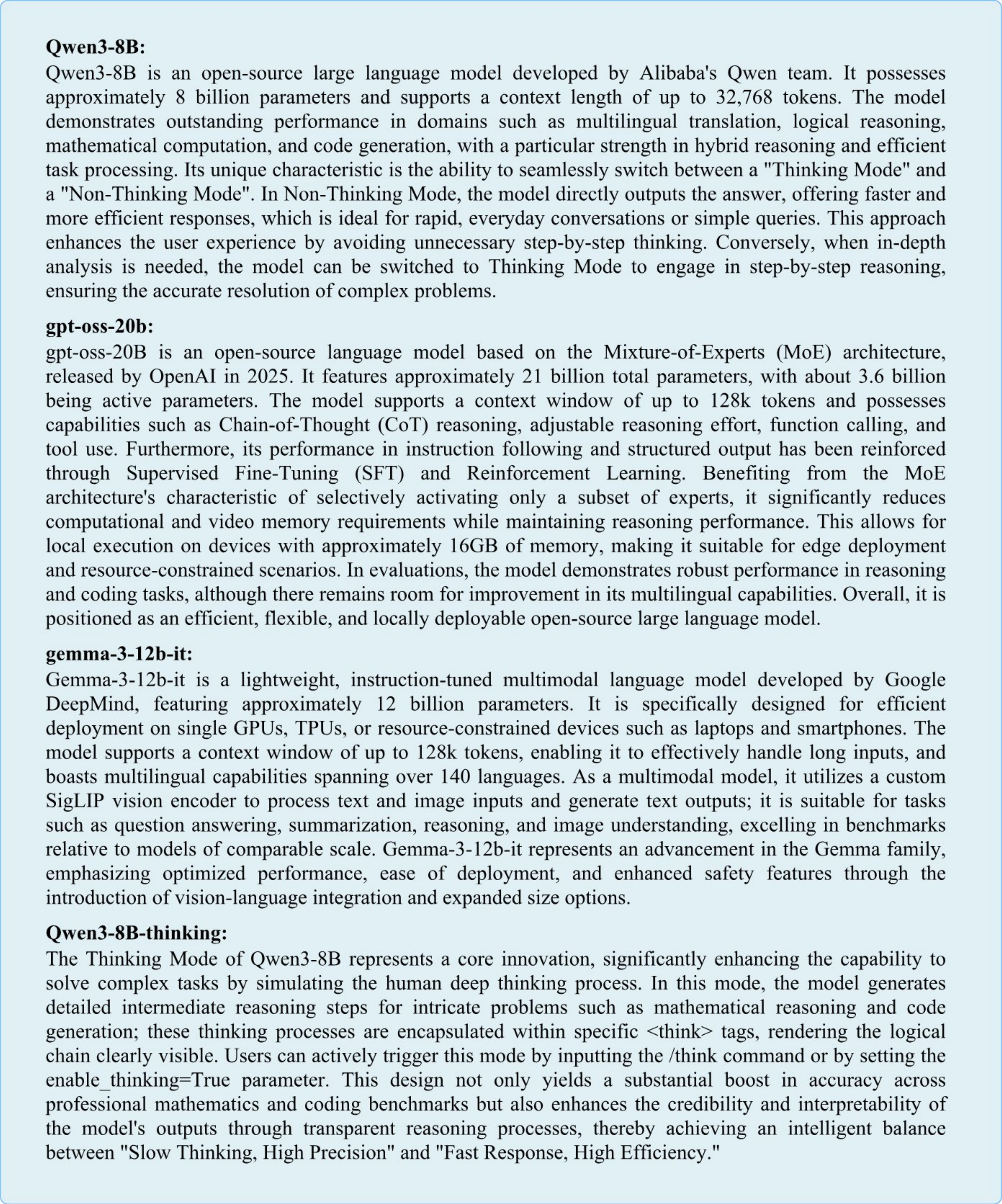}
\end{figure*}

\begin{figure*}[t]
    \centering
    \includegraphics[width=\linewidth]{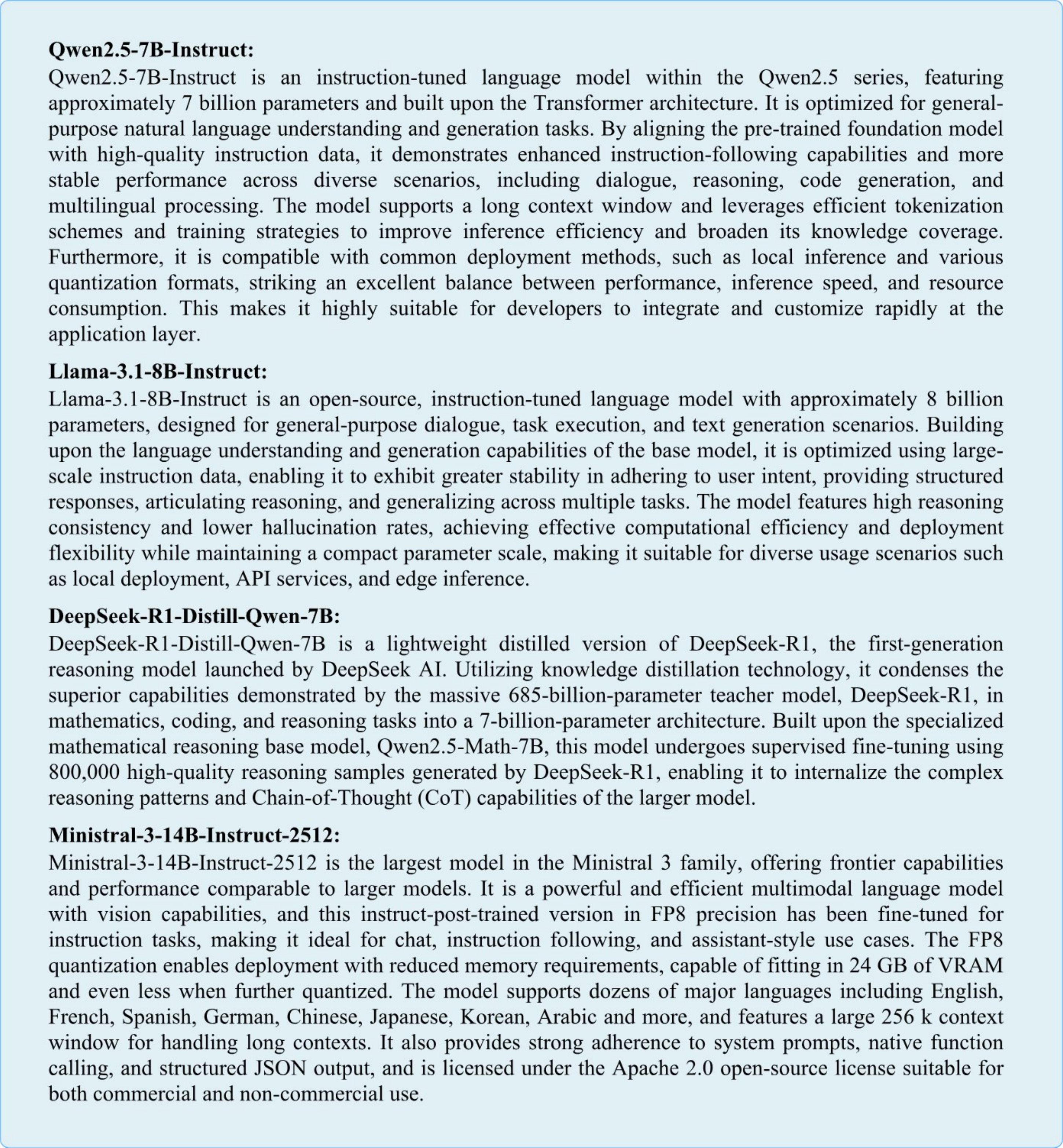}
\end{figure*}

\begin{figure*}[t]
    \centering
    \includegraphics[width=\linewidth]{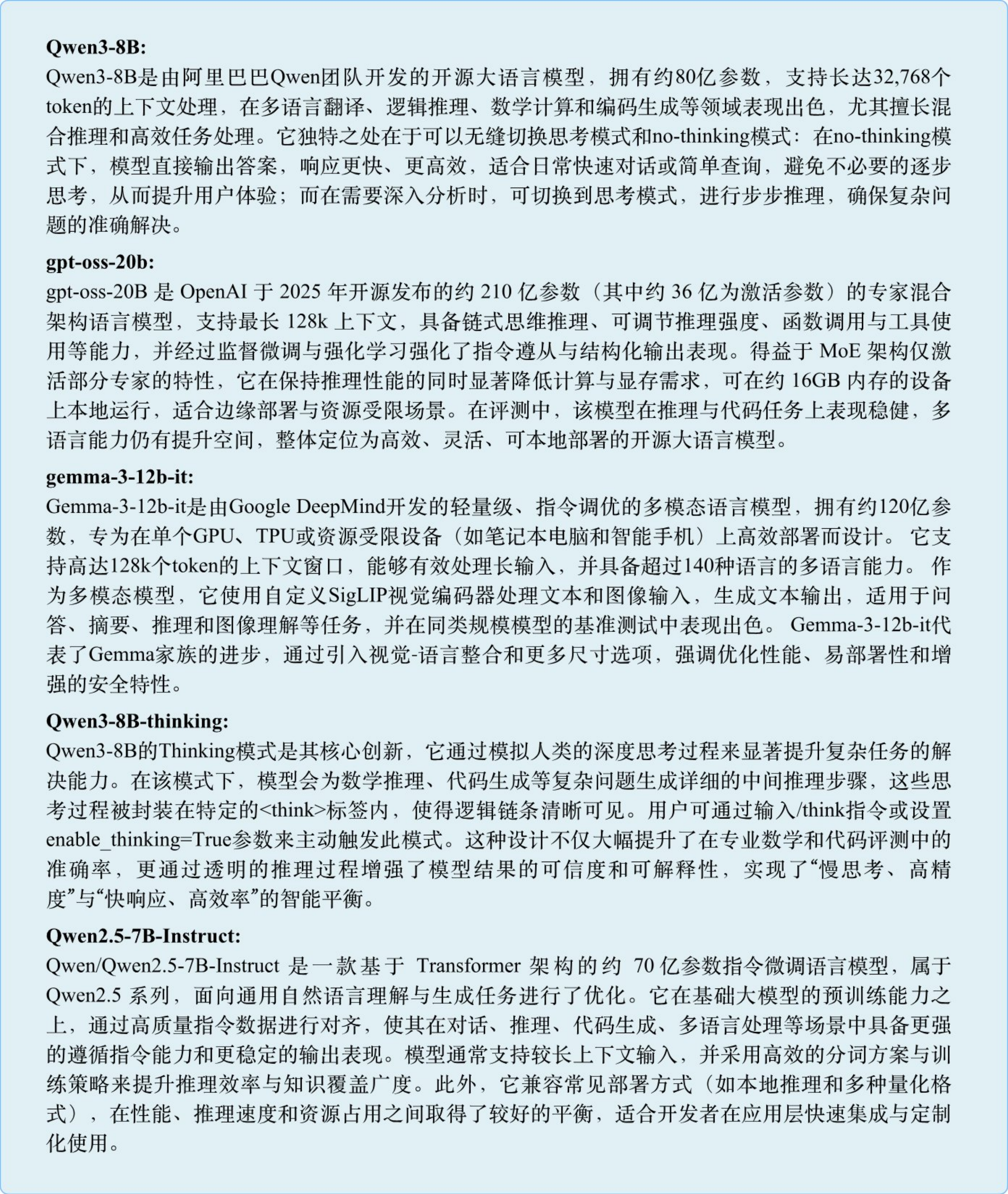}
\end{figure*}

\paragraph{Set Selection.}
The set selection paradigm focuses on selecting and ordering only the positive documents.
Given the full candidate document set as input, the ranker outputs an ordered subset consisting of all positive documents,
with their relative order learned during training and their original retrieval order preserved when ties occur.

\subsection{Answer Extraction and Metric Computation}
\label{app:metric}

All downstream generators are prompted to emit a structured response of the form ``\texttt{Reasoning: ... Answer: ...}''. Given a raw model output, we extract the predicted answer with the following procedure:

\begin{enumerate}[leftmargin=*,itemsep=2pt,topsep=2pt]
    \item \textbf{Marker matching.} If the boldfaced marker \texttt{**Answer:**} appears, we split on its \emph{last} occurrence and take the trailing segment as the predicted answer; otherwise we fall back to the plain \texttt{Answer:} marker, again splitting on the last occurrence. Trailing whitespace is stripped.
    \item \textbf{Parse failure and retries.} If neither marker is present, the response is treated as a parse failure and the generator is re-queried under the temperature-escalation schedule introduced in the main text (temperatures $0.0 \to 0.7 \to 1.0$, up to three attempts).
\end{enumerate}

\paragraph{Normalization.} Following the SQuAD convention, both predictions and ground-truth answers are normalized by (i)~lowercasing, (ii)~removing all ASCII punctuation, (iii)~removing English articles via the regular expression \texttt{\textbackslash{}b(a|an|the)\textbackslash{}b}, and (iv)~collapsing consecutive whitespace into a single space.

\paragraph{Exact Match.} Exact Match is defined as the indicator $\mathbf{1}[\,\mathrm{norm}(\hat{a}) = \mathrm{norm}(a)\,]$, where $\mathrm{norm}(\cdot)$ denotes the normalization above and $\hat{a}$, $a$ are the prediction and ground truth.

\paragraph{F1.} We select the F1 variant according to the language of the ground-truth answers: English references use \textbf{token-level F1}, while Chinese references use \textbf{character-level F1}.
\begin{itemize}[leftmargin=*,itemsep=2pt,topsep=2pt]
    \item \textbf{Token-level F1} (English). After normalization, prediction and ground truth are split on whitespace into token sequences. We compute the multiset intersection via token counts, derive precision and recall against the prediction and reference token sets, and report their harmonic mean.
    \item \textbf{Character-level F1} (Chinese). All whitespace is removed from both strings, and the same multiset-based precision/recall is computed on character sequences. This avoids dependence on a particular Chinese word segmenter.
\end{itemize}

When a query has multiple ground-truth answers, both EM and F1 are computed against each reference and averaged.

\subsection{PRISM Construction Details}
\label{app:prism_details}

\paragraph{Exact dataset statistics.}
PRISM contains $141{,}531$ bilingual queries ($134{,}349$ EN from HotpotQA / 2WikiMultiHopQA / MUSIQUE / MS~MARCO and $7{,}182$ ZH from CRUD-RAG), evaluated against the seven downstream generators in Table~\ref{tab:generators}. On average $\sim$12.6 ordered subsets are probed per (query, generator), yielding $990{,}717$ preference pairs after Stage~\hyperref[sec:stage4_judging]{4} extraction (one per $(q,G)$).

\paragraph{Stage 2 cluster-count distribution.}
Figure~\ref{fig:cluster} reports the empirical distribution of the per-query cluster count $|\mathcal{C}_q|$ induced by the three-axis fusion in Section~\ref{sec:prism_construction}. The mass concentrates in $|\mathcal{C}_q|\in\{4,5,6,7\}$, confirming that the dense-semantic, sparse-lexical, and length axes meaningfully disagree on the majority of candidate pools, so the fused partition does not collapse to a trivial single-cluster regime.

\paragraph{Stage 3(b) template feasibility.}
The six phenomenon-targeted positional templates introduced in Section~\ref{sec:phenomenon} (illustrated in Figure~\ref{fig:schemes}) have minimum positive/negative requirements: the interleaved templates require $n_q^{+}, n_q^{-}\geq 2$, and the head/tail-block and endpoint-anchored templates require $n_q^{+},n_q^{-}\geq 1$. For each Stage~3(a) bag we drop templates whose constraints are not met, so the realized per-query count of ordered subsets can be smaller than the upper bound $K\cdot 6=18$.

\paragraph{Judge configuration.}
\label{app:judge_config}
The LLM-as-a-judge used in Stage~\hyperref[sec:stage4_judging]{4} is \texttt{Gemini-2.5-Flash-Lite}. Each judge call uses the system/user prompts in Figures~\ref{fig:p5}--\ref{fig:p6} and is decoded at temperature $T=0$. If the returned listwise ranking fails schema validation (malformed JSON, missing candidate IDs, or duplicate ranks), the call is retried with \texttt{Gemini-2.5-Flash} as a fallback judge, also at $T=0$.

\paragraph{Stage 4 pair extraction.}
For each $(q,G)$ we form one preference pair by picking the judge's top-ranked subset as $S^{+}_{q,G}$ and the worst-ranked as $S^{-}_{q,G}$, prioritizing the all-positive regime and falling back to the mixed-negative regime for queries that the former does not cover. When the all-positive regime yields an empty dispreferred pool---i.e., all permutations of the positives induce judge-correct answers---we synthesize one dispreferred subset by randomly sampling negatives from $\mathcal{D}_q\setminus\mathcal{P}_q$, so that every $(q,G)$ contributes exactly one pair to $\mathcal{T}$.

\paragraph{Per-generator SFT slice.}
The per-generator SFT slice (Section~\ref{sec:training}) reuses the all-positive $S^{+}_{q,G}$ as its gold output and additionally drops $(q,G)$ pairs whose top-ranked permutation matches the candidates' default sequential order, focusing SFT on non-trivial positional preferences.

\paragraph{Default DPO slice.}
For queries whose top-ranked permutation matches the candidates' default order across all seven generators (no usable preference signal), we build one pair with the canonical positives as $S^{+}$ and $\sim n_q^{+}$ random negatives from $\mathcal{D}_q\setminus\mathcal{P}_q$ as $S^{-}$, conditioned on the generic LLM description.

\subsection{Dataset Sources}
\label{sec:appendix_dataset}

Table~\ref{tab:data_overview} presents a high-level overview of the datasets used for training and evaluation in this work. For clarity and space efficiency, the main table reports only the dataset names, sample sizes, and languages, while additional details regarding data sources, preprocessing, and document augmentation are summarized here.

\paragraph{Training datasets.}
The PRISM\_13K collection consists of multiple widely used open-source question answering benchmarks, including HotpotQA, 2WikiMultiHopQA, MUSIQUE, MS~MARCO, and CRUD-RAG. These datasets originate from heterogeneous sources such as Wikipedia, web search results, and news articles. We apply the dataset-specific filtering and subsampling strategies to control data quality and reasoning depth, while preserving the original question–answer pairs and supervision signals.

\paragraph{Evaluation datasets.}
The evaluation benchmarks cover a diverse set of settings, including web-based question answering (BrowseComp-Plus), knowledge-graph reasoning (KG-MHQA), and news-driven temporal reasoning (ChronoQA), spanning both English and Chinese.

\paragraph{Document augmentation.}
For datasets that are not originally released with explicit retrieval contexts, we construct additional retrieval-augmented generation (RAG) documents to enable document-grounded evaluation. In particular, for SimpleQA and ChineseSimpleQA, we collect relevant documents using an internal search engine pipeline and provide them as external context during inference. This augmentation does not modify the original questions or answers and introduces no additional supervision, but enables a unified RAG-based evaluation setting across all document-augmented benchmarks in this work.

\subsection{Prompt Templates}

\begin{figure*}[t]
    \centering
    \includegraphics[width=\linewidth]{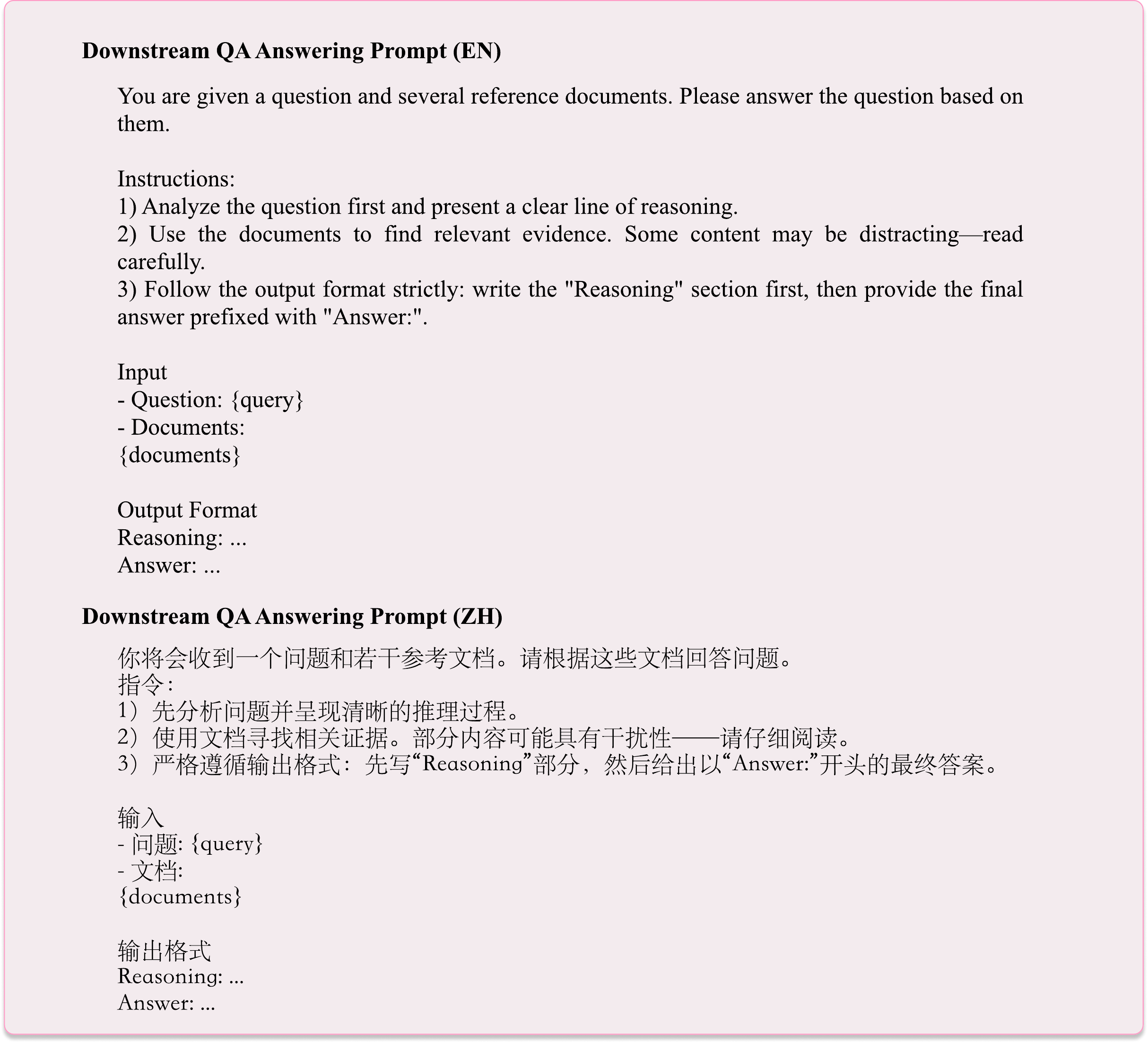}
    \caption{Prompt used for downstream question answering.}
    \label{fig:p1}
\end{figure*}

\begin{figure*}[t]
    \centering
    \includegraphics[width=\linewidth]{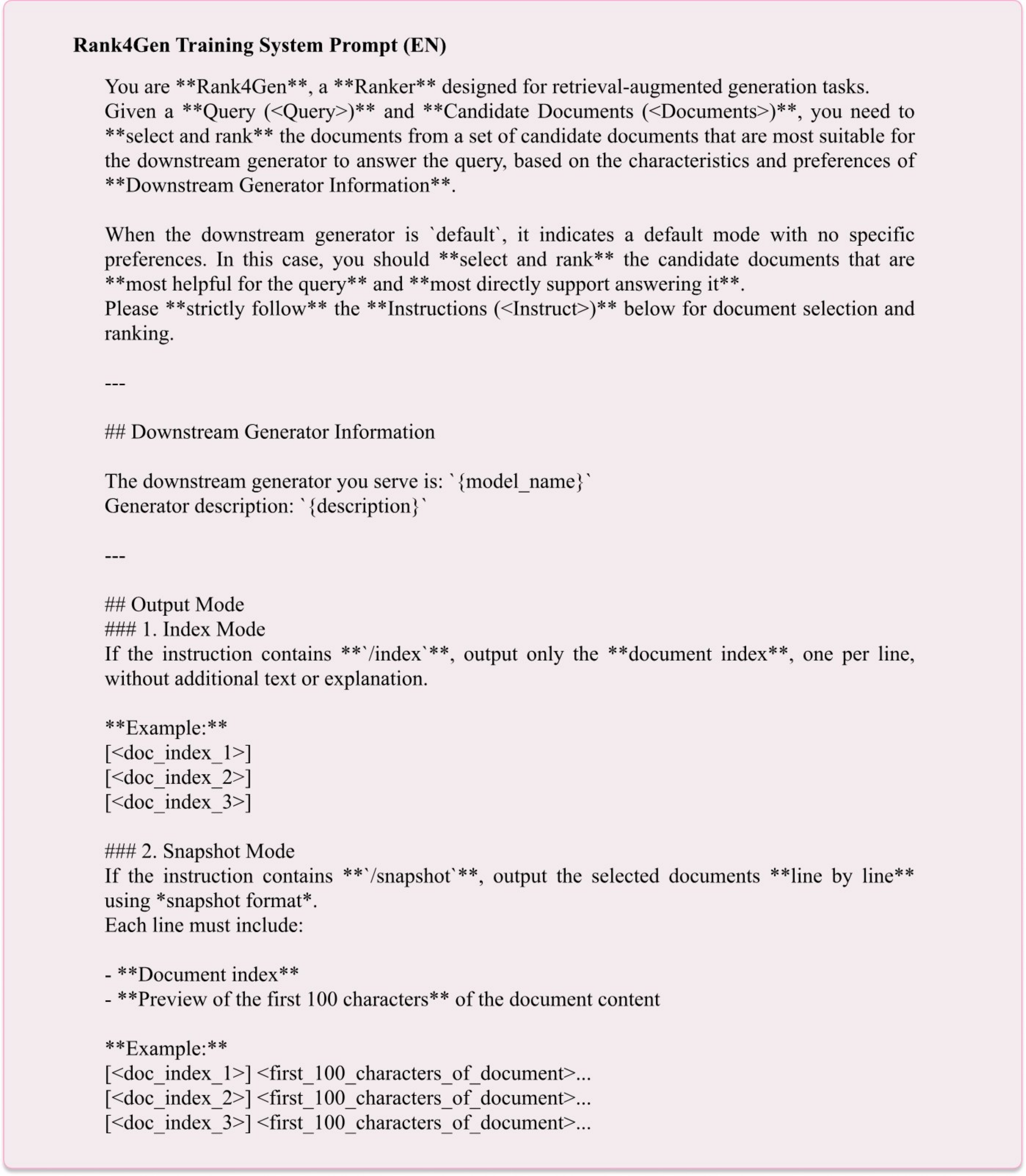}
    \caption{System prompt used for Rank4Gen training (en).}
    \label{fig:p2}
\end{figure*}

\begin{figure*}[t]
    \centering
    \includegraphics[width=\linewidth]{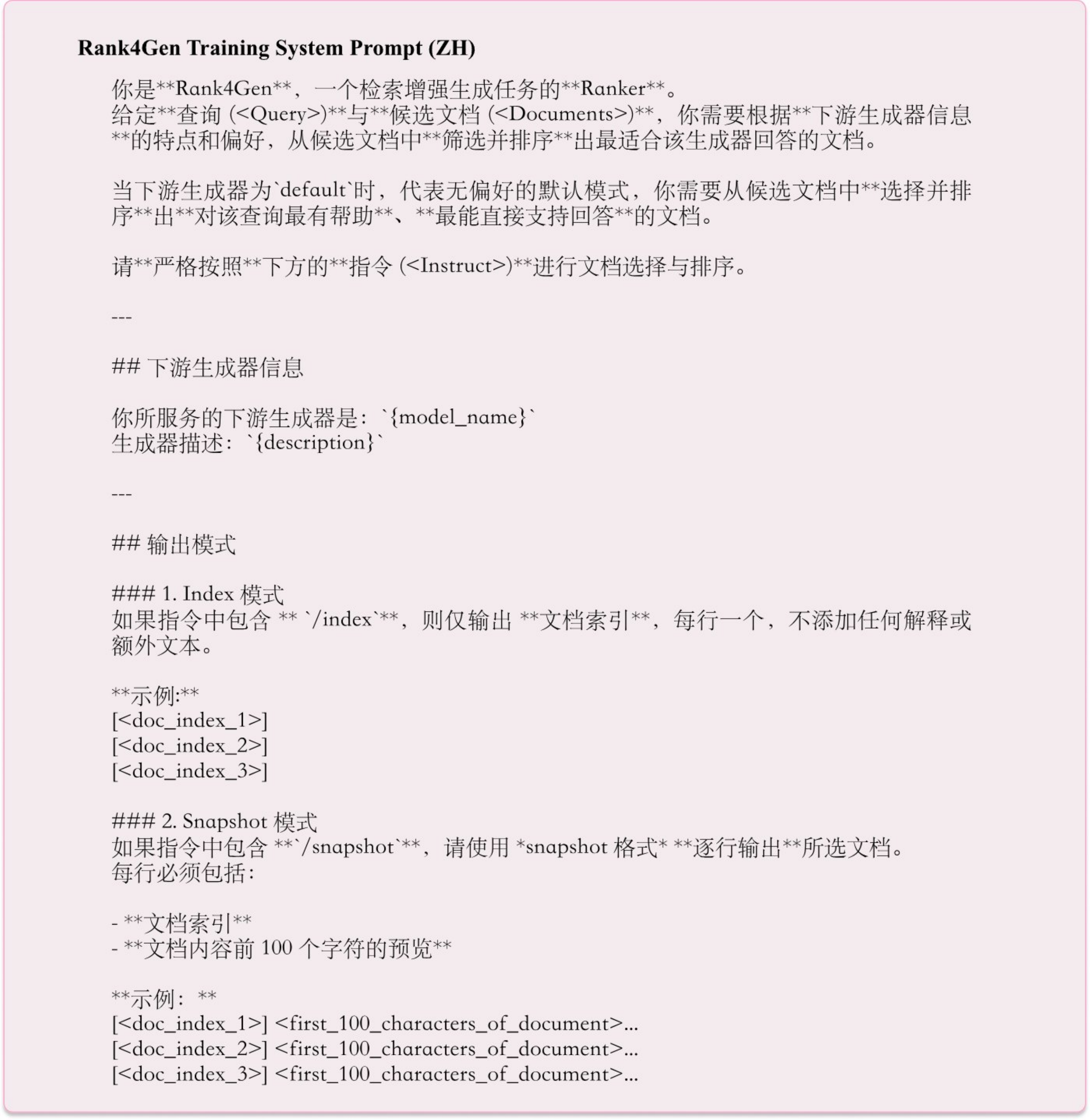}
    \caption{System prompt used for Rank4Gen training (zh).}
    \label{fig:p3}
\end{figure*}

\begin{figure*}[t]
    \centering
    \includegraphics[width=\linewidth]{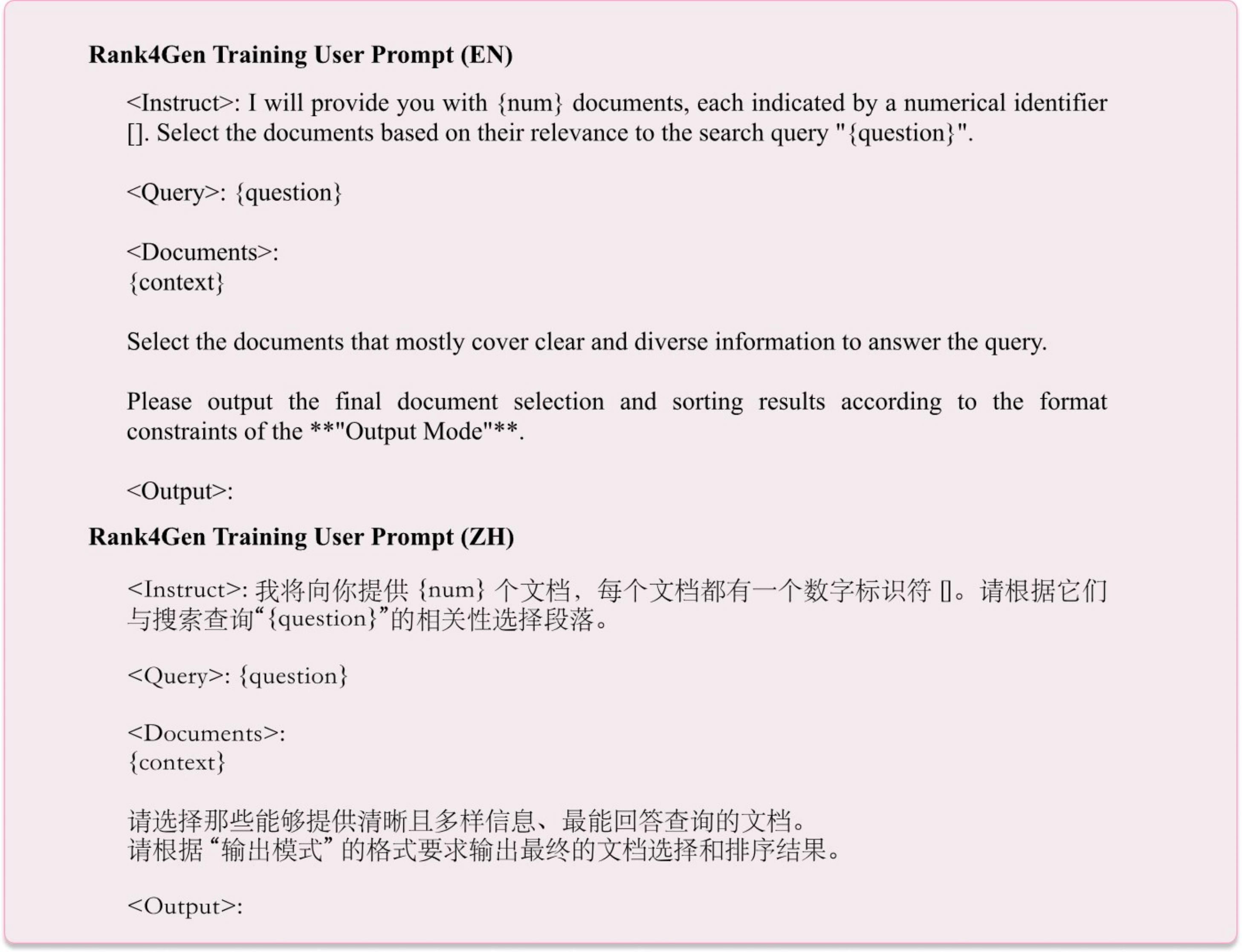}
    \caption{User prompt used for Rank4Gen training.}
    \label{fig:p4}
\end{figure*}

\begin{figure*}[t]
    \centering
    \includegraphics[width=\linewidth]{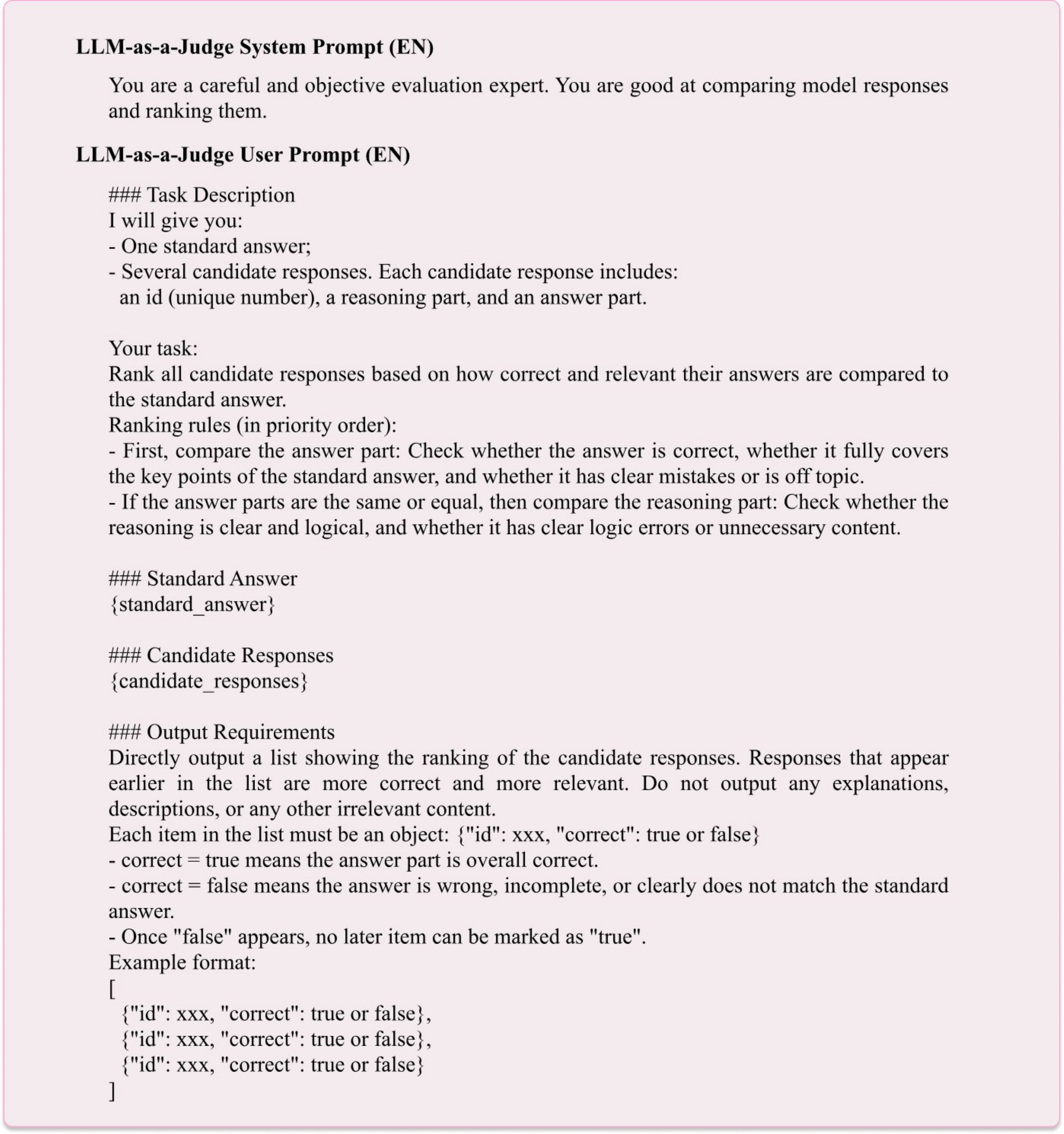}
    \caption{Prompt used for LLM-as-a-Judge (en).}
    \label{fig:p5}
\end{figure*}

\begin{figure*}[t]
    \centering
    \includegraphics[width=\linewidth]{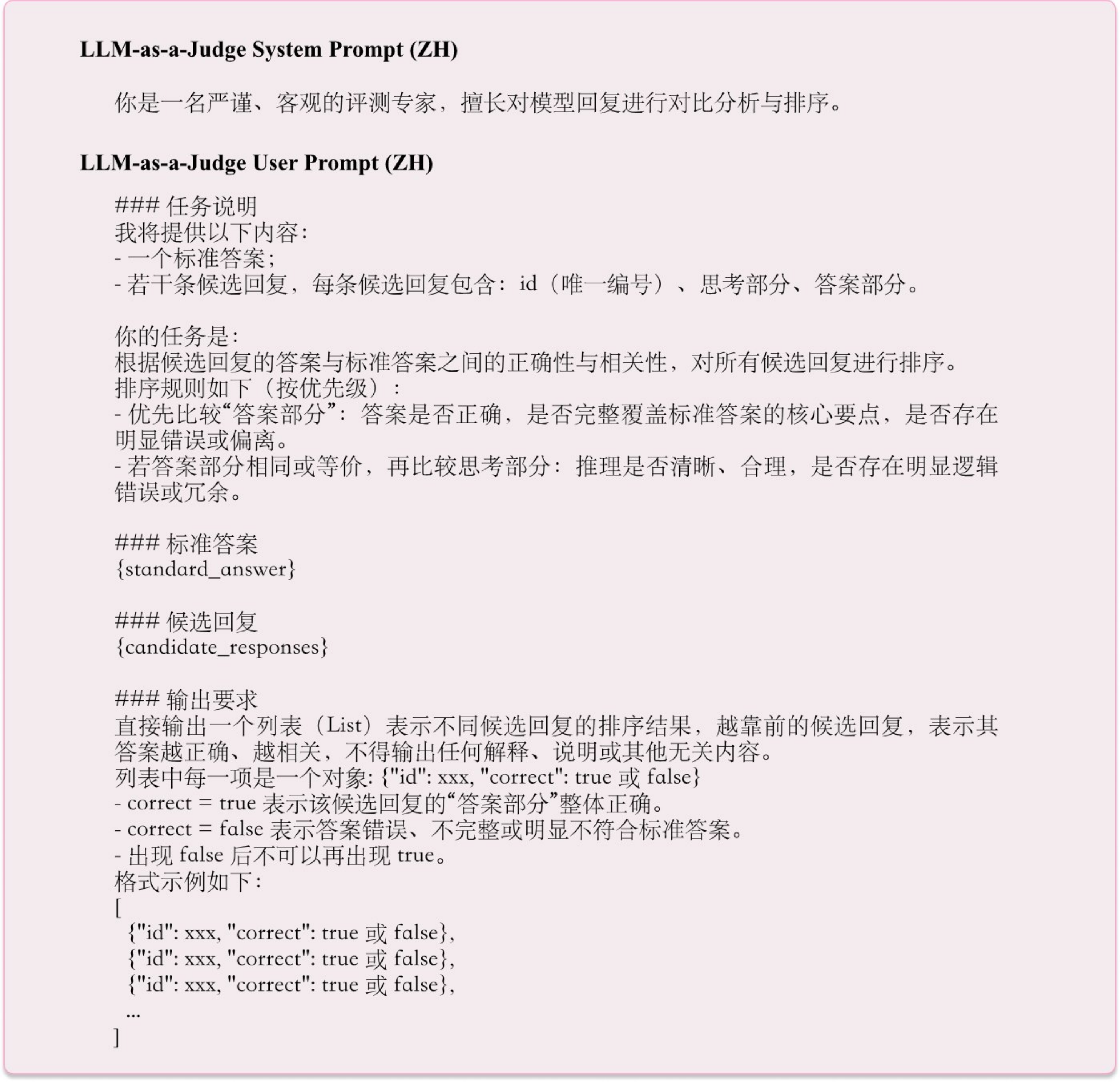}
    \caption{Prompt used for LLM-as-a-Judge (zh).}
    \label{fig:p6}
\end{figure*}

This section presents the prompts used in different stages of our method, as shown in Figures~\ref{fig:p1} to~\ref{fig:p6}.
Figure~\ref{fig:p1} illustrates the prompt template used for downstream question answering in the RAG setting.
Figures~\ref{fig:p2} and~\ref{fig:p3} present the system prompts used for Rank4Gen training in English and Chinese, respectively, while Figure~\ref{fig:p4} shows the corresponding user prompt.
Figures~\ref{fig:p5} and~\ref{fig:p6} show the system and user prompts used for LLM-as-a-Judge during PRISM construction.

\end{document}

%% file: tables/Data_Collection.tex
\begin{table}[t]
    \centering
    \small
    \resizebox{1.0\linewidth}{!}{
    \begin{tabular}{l|ccc}
        \toprule
        \textbf{Dataset} 
        & \textbf{Samples} 
        & \textbf{Language} \\
        \midrule
        
        \multicolumn{3}{c}{\textbf{PRISM\_13K}} \\
        \midrule
        
        HotpotQA~\citep{yang-etal-2018-hotpotqa}
        & 4,000
        & EN \\
        
        2WikiMultiHopQA~\citep{ho-etal-2020-constructing}
        & 2,000
        & EN \\
        
        MUSIQUE~\citep{trivedi-etal-2022-musique}
        & 2,000
        & EN \\
        
        MS MARCO~\citep{nguyen2016ms}
        & 2,000
        & EN \\
        
        CRUD-RAG~\citep{lyu2025crud}
        & 2,994
        & ZH \\
        
        \midrule
        \multicolumn{3}{c}{\textbf{Evaluation Datasets}} \\
        \midrule
        
        BrowseComp-Plus~\citep{chen2025browsecomp}
        & 830
        & EN \\
        
        KG-MHQA~\citep{wang2025kg}
        & 307
        & EN \\
        
        ChronoQA~\citep{chen2025question}
        & 5,173
        & ZH \\
        
        SimpleQA~\citep{wei2024measuring}
        & 4,326
        & EN \\
        
        ChineseSimpleQA~\citep{he-etal-2025-chinese}
        & 3,000
        & ZH \\
        
        \bottomrule
    \end{tabular}}
    \caption{Overview of training and evaluation datasets.
    Additional details on data collection and construction are 
    provided in the Section~\ref{sec:prism_construction} and Appendix~\ref{sec:appendix_dataset}.}
    \label{tab:data_overview}
\end{table}

%% file: tables/OOD.tex
\definecolor{skyblue}{RGB}{135, 206, 235}

\begin{table*}[ht]
    \centering
    \small
    \resizebox{1.0\linewidth}{!}{
    \begin{tabular}{lc|cc|cc|cc|cc|cc|cc}
        \toprule
        \multirow{2}{*}{\textbf{Ranker}} &
        \multirow{2}{*}{\textbf{Generator}} &
        \multicolumn{2}{c|}{\textbf{BrowseComp+}} &
        \multicolumn{2}{c|}{\textbf{KG-MHQA}} &
        \multicolumn{2}{c|}{\textbf{ChronoQA}} &
        \multicolumn{2}{c|}{\textbf{SimpleQA}} &
        \multicolumn{2}{c|}{\textbf{CN-SimpleQA}} &
        \multicolumn{2}{c}{\textbf{Avg.}}\\
        & &
        EM & F1 &
        EM & F1 &
        EM & F1 &
        EM & F1 &
        EM & F1 &
        EM & F1 \\
        \midrule
        \multicolumn{14}{c}{\textbf{Without RAG}} \\
        \midrule
        \multirow{4}{*}{\textbf{-}}
            & Qwen3-8B & 0.12 & 3.53 & 0.65 & 3.36 & 2.84 & 24.18 & 2.80 & 9.73 & 14.67 & 46.27 & 4.22 & 17.41 \\
            & gemma-3-12b-it & 0.00 & 2.45 & 0.65 & 3.21 & 1.70 & 13.66 & 4.28 & 11.54 & 9.07 & 34.28 & 3.14 & 13.03 \\
            & Llama-3.1-8B-Instruct & 0.00 & 1.50 & 0.65 & 2.76 & 2.01 & 20.77 & 4.05 & 9.45 & 6.30 & 30.61 & 2.60 & 13.02 \\
            & DeepSeek-R1-Distill-Qwen-7B & 0.12 & 1.62 & 0.33 & 2.48 & 1.45 & 21.80 & 1.16 & 5.16 & 2.50 & 26.69 & 1.11 & 11.55 \\
        \midrule
        \multicolumn{14}{c}{\textbf{Embedding}} \\
        \midrule
        \multirow{4}{*}{\textbf{bge-m3}}
            & Qwen3-8B & 17.71 & 23.03 & 5.54 & 8.89 & 7.85 & 32.73 & 60.80 & 75.55 & 31.87 & 48.71 & 24.75 & 37.78 \\
            & gemma-3-12b-it & 15.54 & 19.82 & 5.54 & 8.52 & 2.73 & 15.05 & 61.30 & 76.28 & 25.83 & 44.37 & 22.19 & 32.81 \\
            & Llama-3.1-8B-Instruct & 10.72 & 15.75 & 4.56 & 6.95 & 4.60 & 30.38 & 59.02 & 73.87 & 15.33 & 36.60 & 18.85 & 32.71 \\
            & DeepSeek-R1-Distill-Qwen-7B & 2.41 & 5.37 & 3.58 & 5.98 & 0.48 & 17.01 & 13.64 & 35.63 & 18.07 & 35.60 & 7.64 & 19.92 \\
        \midrule
        \multicolumn{14}{c}{\textbf{Pointwise}$^{\tau}$} \\
        \midrule
        \multirow{4}{*}{\textbf{Pointwise-Vanilla}}
            & Qwen3-8B & \underline{46.02} & \underline{55.20} & 6.19 & 11.40 & \underline{9.39} & \underline{36.83} & 60.56 & 75.75 & 30.83 & 48.09 & 30.60 & 45.45 \\
            & gemma-3-12b-it & 39.44 & 46.64 & 6.19 & 11.23 & 3.15 & 16.77 & 60.61 & 76.30 & 24.83 & 42.94 & 26.84 & 38.78 \\
            & Llama-3.1-8B-Instruct & 33.38 & 42.47 & 5.86 & 9.08 & \underline{5.39} & 33.52 & 58.34 & 73.66 & 15.80 & 36.60 & 23.75 & 39.07 \\
            & DeepSeek-R1-Distill-Qwen-7B & 8.31 & 15.47 & 5.54 & 9.30 & 0.81 & 21.14 & 14.24 & 35.55 & 17.07 & 34.10 & 9.19 & 23.11 \\
        \midrule
        \multicolumn{14}{c}{\textbf{Listwise}$^{\tau}$} \\
        \midrule
        \multirow{4}{*}{\textbf{Listwise-Vanilla}}
            & Qwen3-8B & 29.88 & 37.77 & \underline{7.82} & 11.52 & 8.25 & 34.13 & 59.69 & 74.79 & 32.60 & 49.40 & 27.65 & 41.52 \\
            & gemma-3-12b-it & 25.42 & 31.08 & \underline{7.82} & \underline{12.20} & 3.17 & 16.12 & 60.33 & 75.65 & 27.03 & 45.11 & 24.75 & 36.03 \\
            & Llama-3.1-8B-Instruct & 23.01 & 30.25 & 6.51 & 9.76 & 5.20 & 32.11 & 58.07 & 73.56 & 17.03 & 38.80 & 21.96 & 36.90 \\
            & DeepSeek-R1-Distill-Qwen-7B & 4.46 & 8.92 & 6.19 & 9.80 & 0.83 & 19.00 & 13.71 & 35.60 & 19.07 & 37.10 & 8.85 & 22.08 \\
        \midrule
        \multirow{4}{*}{\textbf{RankZephyr}}
            & Qwen3-8B & 45.06 & 53.74 & 7.49 & 12.09 & 9.09 & 35.75 & \textbf{61.35} & \textbf{76.38} & 31.60 & 48.83 & 30.92 & 45.36 \\
            & gemma-3-12b-it & 41.57 & 48.90 & 7.17 & 11.23 & 3.02 & 15.89 & \textbf{61.65} & \textbf{76.95} & 25.83 & 44.45 & 27.85 & 39.48 \\
            & Llama-3.1-8B-Instruct & 34.58 & 43.71 & 6.51 & 9.30 & 5.06 & 33.26 & 58.00 & 73.98 & 14.53 & 36.74 & 23.74 & 39.40 \\
            & DeepSeek-R1-Distill-Qwen-7B & 7.71 & 14.45 & 6.19 & 9.24 & 0.58 & 18.48 & 13.80 & 36.46 & 17.43 & 35.31 & 9.14 & 22.79 \\
        \midrule
        \multicolumn{14}{c}{\textbf{Set Selection}} \\
        \midrule        
            \rowcolor{skyblue!8} & Qwen3-8B & 45.54 & 53.87 & \textbf{8.14} & \underline{12.76} & 8.89 & 35.05 & \underline{61.03} & \underline{75.76} & \underline{35.77} & \underline{51.58} & \underline{31.87} & \underline{45.80} \\
            \rowcolor{skyblue!8} & gemma-3-12b-it & \underline{46.63} & \underline{53.52} & 7.49 & 12.09 & 4.52 & 18.28 & \underline{61.47} & \underline{76.82} & \underline{29.40} & \underline{46.97} & \underline{29.90} & \underline{41.54} \\
            \rowcolor{skyblue!8} & Llama-3.1-8B-Instruct & \underline{39.76} & \underline{48.77} & \underline{6.84} & \underline{10.46} & 4.68 & \underline{35.68} & \underline{59.96} & \underline{75.34} & \underline{33.10} & \underline{56.69} & \underline{28.87} & \underline{45.39} \\
            \rowcolor{skyblue!8}
        \multirow{-4}{*}{\textbf{SetSelection-Vanilla}}
            & DeepSeek-R1-Distill-Qwen-7B & \underline{15.90} & \underline{25.06} & \underline{6.84} & \underline{11.18} & \textbf{8.14} & \underline{34.62} & \textbf{36.04} & \underline{55.23} & \textbf{33.27} & \underline{50.31} & \underline{20.04} & \underline{35.28} \\
        \midrule
        \multirow{4}{*}{\textbf{SETR}}
            & Qwen3-8B & 29.64 & 33.71 & 5.86 & 9.10 & \textbf{9.59} & \textbf{36.89} & 53.95 & 67.42 & 32.83 & 49.58 & 26.37 & 39.34 \\
            & gemma-3-12b-it & 28.67 & 32.41 & 5.86 & 9.55 & \underline{4.62} & \underline{19.55} & 54.14 & 68.36 & 26.67 & 44.97 & 23.99 & 34.97 \\
            & Llama-3.1-8B-Instruct & 25.06 & 30.33 & 5.54 & 8.64 & 5.08 & 35.43 & 53.49 & 67.51 & 23.47 & 46.16 & 22.53 & 37.61 \\
            & DeepSeek-R1-Distill-Qwen-7B & 11.57 & 17.83 & 4.56 & 8.23 & 6.19 & 31.70 & 31.16 & 48.83 & 27.07 & 43.65 & 16.11 & 30.05 \\
        \midrule
            \rowcolor{blue!8} & Qwen3-8B & \textbf{46.51} & \textbf{55.91} & \textbf{8.14} & \textbf{14.12} & 9.28 & 36.23 & 59.80 & 75.26 & \textbf{35.83} & \textbf{52.84} & \textbf{31.91} & \textbf{46.87} \\
            \rowcolor{blue!8} & gemma-3-12b-it & \textbf{46.87} & \textbf{54.88} & \textbf{8.47} & \textbf{14.01} & \textbf{5.43} & \textbf{20.64} & 61.35 & 76.31 & \textbf{31.70} & \textbf{50.64} & \textbf{30.76} & \textbf{43.30} \\
            \rowcolor{blue!8} & Llama-3.1-8B-Instruct & \textbf{42.05} & \textbf{52.86} & \textbf{7.49} & \textbf{12.05} & \textbf{5.74} & \textbf{37.66} & \textbf{60.10} & \textbf{76.09} & \textbf{33.47} & \textbf{58.21} & \textbf{29.77} & \textbf{47.37} \\
        \rowcolor{blue!8}
        \multirow{-4}{*}{\textbf{Rank4Gen}}
            & DeepSeek-R1-Distill-Qwen-7B & \textbf{18.43} & \textbf{31.14} & \textbf{8.14} & \textbf{13.06} & \underline{7.29} & \textbf{35.62} & \underline{34.91} & \textbf{55.80} & \underline{33.17} & \textbf{51.16} & \textbf{20.39} & \textbf{37.36} \\
        \bottomrule
    \end{tabular}}
    \caption{Main results on five RAG benchmarks using different rankers and representative generators. Performance is reported in Exact Match (EM) and token-level F1. $\tau$ denotes that the top-10 retrieved documents are used. The best and second-best results are highlighted in \textbf{bold} and \underline{underline} for each generator, respectively.}
    \label{tab:main_results}
\end{table*}

%% file: tables/Ablation_table.tex
\begin{table*}[ht]
    \centering
    \small
    \resizebox{1.0\linewidth}{!}{
    \begin{tabular}{lcc|cc|cc|cc|cc|cc|cc}
        \toprule
        \multirow{2}{*}{\textbf{Training Setting}} &
        \multirow{2}{*}{\textbf{Infer. Mode}} &
        \multirow{2}{*}{\textbf{Generator}} &
        \multicolumn{2}{c|}{\textbf{BrowseComp+}} &
        \multicolumn{2}{c|}{\textbf{KG-MHQA}} &
        \multicolumn{2}{c|}{\textbf{ChronoQA}} &
        \multicolumn{2}{c|}{\textbf{SimpleQA}} &
        \multicolumn{2}{c|}{\textbf{CN-SimpleQA}} &
        \multicolumn{2}{c}{\textbf{Avg.}} \\
        
        & \multicolumn{2}{c|}{}  &
        EM & F1 &
        EM & F1 &
        EM & F1 &
        EM & F1 &
        EM & F1 &
        EM & F1 \\
        \midrule
        
        \multirow{4}{*}{
              \makecell[l]{
                Qwen3-8B \\
                {\footnotesize\emph{Zero-shot}}
              }
            }
            & \multirow{4}{*}{/index} 
            & Qwen3-8B & 33.98 & 41.05 & 7.17 & 12.01 & 8.29 & 34.80 & \textbf{61.78} & \textbf{76.60} & 32.30 & 49.08 & 28.70 & 42.71 \\
            & & gemma-3-12b-it & 21.93 & 26.03 & 7.49 & 11.91 & 3.07 & 15.17 & \textbf{62.00} & \textbf{77.49} & 25.17 & 42.70 & 23.93 & 34.66 \\
            & & Llama-3.1-8B-Instruct & 17.71 & 23.70 & 5.86 & 9.05 & 5.24 & 32.81 & 58.37 & 73.94 & 15.07 & 36.65 & 20.45 & 35.23 \\
            & & DeepSeek-R1-Distill-Qwen-7B & 4.58 & 9.45 & 4.56 & 8.22 & 2.30 & 21.30 & 19.19 & 39.84 & 17.77 & 34.59 & 9.68 & 22.68 \\
        \midrule
        
        \multirow{8}{*}{SFT+DPO} 
            & \multirow{4}{*}{/snapshot} 
            & Qwen3-8B & \textbf{46.51} & \textbf{55.91} & \textbf{8.14} & \textbf{14.12} & \textbf{9.28} & \textbf{36.23} & \underline{59.80} & \underline{75.26} & \textbf{35.83} & \textbf{52.84} & \textbf{31.91} & \textbf{46.87} \\
            & & gemma-3-12b-it & \textbf{46.87} & \textbf{54.88} & \underline{8.47} & \underline{14.01} & \textbf{5.43} & \underline{20.64} & 61.35 & 76.31 & \textbf{31.70} & \textbf{50.64} & \textbf{30.76} & \textbf{43.30} \\
            & & Llama-3.1-8B-Instruct & \textbf{42.05} & \textbf{52.86} & \textbf{7.49} & \textbf{12.05} & \underline{5.74} & \underline{37.66} & \textbf{60.10} & \textbf{76.09} & \textbf{33.47} & \textbf{58.21} & \textbf{29.77} & \textbf{47.37} \\
            & & DeepSeek-R1-Distill-Qwen-7B & \underline{18.43} & \textbf{31.14} & \textbf{8.14 }& \textbf{13.06} & \textbf{7.29} & \textbf{35.62} & \textbf{34.91} & \textbf{55.80} & \underline{33.17} & \underline{51.16} & \textbf{20.39} & \textbf{37.36} \\
            & \multirow{4}{*}{/index}
            & Qwen3-8B & 45.78 & 54.51 & \textbf{8.14} & \underline{14.04} & \textbf{9.28} & \underline{36.15} & 57.51 & 72.73 & 33.40 & 50.77 & 30.82 & \underline{45.64} \\
            & & gemma-3-12b-it & 44.10 & 52.45 & \textbf{8.79} & \textbf{14.19} & \underline{5.16} & \textbf{22.20} & 59.04 & 74.14 & \textbf{31.70} & \underline{50.50} & 29.76 & \underline{42.70} \\
            & & Llama-3.1-8B-Instruct & 38.67 & 48.83 & 6.84 & 11.22 & \textbf{5.78} & \textbf{37.67} & 58.88 & 74.61 & 26.87 & 53.13 & 27.41 & 45.09 \\
            & & DeepSeek-R1-Distill-Qwen-7B & \textbf{18.67} & \underline{29.55} & \underline{7.17} & \underline{12.46} & 4.68 & 32.40 & 33.22 & 53.39 & 29.90 & 47.81 & 18.73 & 35.12 \\
         \multirow{4}{*}{SFT} 
            & \multirow{4}{*}{/snapshot}
            & Qwen3-8B & 44.34 & 52.91 & \textbf{8.14} & 13.23 & \underline{8.87} & 35.55 & 56.93 & 71.89 & \underline{34.10} & 50.52 & 30.48 & 44.82 \\
            & & gemma-3-12b-it & \underline{45.42} & \underline{53.70} & 7.49 & 13.08 & 4.37 & 18.30 & 57.63 & 72.99 & 29.77 & 48.40 & 28.94 & 41.29 \\
            & & Llama-3.1-8B-Instruct & 39.16 & 49.26 & \underline{7.17} & \underline{11.83} & 4.37 & 34.89 & 57.88 & 73.65 & \underline{31.50} & \underline{56.09} & \underline{28.02} & \underline{45.14} \\
            & & DeepSeek-R1-Distill-Qwen-7B & 17.23 & 28.81 & 6.84 & 12.04 & \underline{7.85} & \underline{34.30} & 33.47 & 53.37 & \textbf{33.67} & \textbf{51.23} & \underline{19.81} & \underline{35.95} \\
         \multirow{4}{*}{
            \makecell[l]{
                SFT+DPO \\
                {\footnotesize\emph{Index-only}}
              }
        } 
            & \multirow{4}{*}{/index}
            & Qwen3-8B & \underline{46.02} & \underline{54.47} & \underline{7.82} & 13.33 & 8.83 & 34.60 & 59.45 & 74.56 & 34.07 & \underline{50.83} & \underline{31.24} & 45.56 \\
            & & gemma-3-12b-it & 44.22 & 51.97 & 7.82 & 12.96 & 4.72 & 20.36 & \underline{61.44} & \underline{76.35} & \underline{31.60} & 50.45 & \underline{29.96} & 42.42 \\
            & & Llama-3.1-8B-Instruct & \underline{39.76} & \underline{50.00} & \underline{7.17} & 11.72 & 4.89 & 34.61 & \underline{59.75} & \underline{76.03} & 26.87 & 51.98 & 27.69 & 44.87 \\
            & & DeepSeek-R1-Distill-Qwen-7B & 14.10 & 26.38 & \underline{7.17} & 11.76 & 4.85 & 31.42 & \underline{33.77} & \underline{54.22} & 30.20 & 47.20 & 18.02 & 34.20 \\
        \bottomrule
    \end{tabular}}
    \caption{Ablation study results with different training settings and inference modes for Rank4Gen. The best and second-best results are highlighted in \textbf{bold} and \underline{underline} for each generator, respectively.}
    \label{tab:ablation_rank4gen}
\end{table*}

%% file: tables/Generalised_Explanation.tex
\begin{table}[ht]
    \centering
    \small
    \resizebox{1.0\linewidth}{!}{
    \begin{tabular}{lc|cc|cc}
        \toprule
        \textbf{Generator} & \textbf{Setting} &
        \multicolumn{2}{c}{\textbf{BrowseComp+}} &
        \multicolumn{2}{c}{\textbf{ChronoQA}} \\
        & & EM & F1 & EM & F1 \\
        \midrule

        \multirow{3}{*}{Ministral-3-14B}
        & \textbf{w/o RAG}
        & 0.24 & 2.26
        & 0.65 & 2.17 \\

        & \textbf{Rank4Gen} {\footnotesize\emph{default}}
        & 35.78 & 49.04
        & 0.85 & 21.78 \\

        & \textbf{Rank4Gen}
        & 36.27 & 50.89
        & 0.93 & 22.23 \\

        \midrule

        \multirow{3}{*}{DeepSeek-V3.2}
        & \textbf{w/o RAG}
        & 2.41 & 5.41
        & 1.30 & 3.79 \\

        & \textbf{Rank4Gen} {\footnotesize\emph{default}}
        & 52.53 & 63.04
        & 7.60 & 33.37 \\

        & \textbf{Rank4Gen}
        & 53.73 & 63.04
        & 7.97 & 33.56 \\

        \bottomrule
    \end{tabular}}
    \caption{Generalization results on BrowseComp+ and ChronoQA with two OOD generators (EM / F1).}
    \label{tab:generalization}
\end{table}

%% file: tables/generators.tex
\begin{table}[htbp]
    \centering \small
    \begin{tabular}{c}
        \toprule
        \rowcolor{gray!20}\textbf{Generators within PRISM}  \\
        \midrule
        
        Qwen3-8B \\
        
        gpt-oss-20b \\
        
        gemma-3-12b-it \\
        
        Qwen3-8B-thinking \\
        
        Qwen2.5-7B-Instruct \\

        Llama-3.1-8B-Instruct \\
        
        DeepSeek-R1-Distill-Qwen-7B \\
        
        \midrule
        \rowcolor{gray!20}\textbf{Generators without PRISM} \\
        \midrule
        
        Ministral-3-14B-Instruct-2512 \\
        
        DeepSeek-V3.2 \\
        
        \bottomrule
    \end{tabular}
    \caption{Generators within PRISM and without PRISM.}
    \label{tab:generators}
\end{table}

%% file: tables/appendex_table.tex
\begin{table*}[ht]
    \centering
    \small
    \resizebox{1.0\linewidth}{!}{
    \begin{tabular}{lc|cc|cc|cc|cc|cc|cc}
        \toprule
        \multirow{2}{*}{\textbf{Ranker}} &
        \multirow{2}{*}{\textbf{Generator}} &
        \multicolumn{2}{c|}{\textbf{BrowseComp+}} &
        \multicolumn{2}{c|}{\textbf{KG-MHQA}} &
        \multicolumn{2}{c|}{\textbf{ChronoQA}} &
        \multicolumn{2}{c|}{\textbf{SimpleQA}} &
        \multicolumn{2}{c|}{\textbf{CN-SimpleQA}} &
        \multicolumn{2}{c}{\textbf{Avg.}} \\
        & &
        EM & F1 &
        EM & F1 &
        EM & F1 &
        EM & F1 &
        EM & F1 &
        EM & F1 \\
        \midrule
        \multicolumn{14}{c}{\textbf{Without RAG}} \\
        \midrule
        \multirow{7}{*}{\textbf{-}}
            & Qwen3-8B & 0.12 & 3.53 & 0.65 & 3.36 & 2.84 & 24.18 & 2.80 & 9.73 & 14.67 & 46.27 & 4.22 & 17.41 \\
            & gpt-oss-20b & 0.24 & 0.83 & 0.98 & 2.79 & 1.72 & 27.78 & 3.26 & 11.11 & 6.87 & 37.48 & 2.61 & 16.00 \\
            & gemma-3-12b-it & 0.00 & 2.45 & 0.65 & 3.21 & 1.70 & 13.66 & 4.28 & 11.54 & 9.07 & 34.28 & 3.14 & 13.03 \\
            & Qwen3-8B-thinking & 0.00 & 1.14 & 0.65 & 2.96 & 2.07 & 22.21 & 3.17 & 9.99 & 14.47 & 44.33 & 4.07 & 16.13 \\
            & Qwen2.5-7B-Instruct & 0.12 & 2.54 & 0.65 & 2.82 & 0.56 & 13.43 & 1.04 & 4.34 & 0.90 & 16.58 & 0.65 & 7.94 \\
            & Llama-3.1-8B-Instruct & 0.00 & 1.50 & 0.65 & 2.76 & 2.01 & 20.77 & 4.05 & 9.45 & 6.30 & 30.61 & 2.60 & 13.02 \\
            & DeepSeek-R1-Distill-Qwen-7B & 0.12 & 1.62 & 0.33 & 2.48 & 1.45 & 21.80 & 1.16 & 5.16 & 2.50 & 26.69 & 1.11 & 11.55 \\
        \midrule
        \multicolumn{14}{c}{\textbf{Pointwise}$^{\tau}$} \\
        \midrule
        \multirow{7}{*}{\textbf{Pointwise-Vanilla}}
            & Qwen3-8B & \underline{46.02} & \underline{55.20} & 6.19 & 11.40 & \underline{9.39} & \underline{36.83} & 60.56 & 75.75 & 30.83 & 48.09 & 30.60 & 45.45 \\
            & gpt-oss-20b & 33.13 & 43.39 & 6.19 & 8.66 & 2.26 & 18.38 & 41.22 & 59.89 & 19.20 & 43.23 & 20.40 & 34.71 \\
            & gemma-3-12b-it & 39.44 & 46.64 & 6.19 & 11.23 & 3.15 & 16.77 & 60.61 & 76.30 & 24.83 & 42.94 & 26.84 & 38.78 \\
            & Qwen3-8B-thinking & \textbf{49.28} & \underline{56.80} & 7.82 & \underline{12.57} & 8.58 & 34.02 & 58.88 & 74.22 & 27.40 & 45.13 & 30.39 & 44.55 \\
            & Qwen2.5-7B-Instruct & 12.53 & 30.37 & 4.89 & 9.30 & 4.21 & \textbf{30.95} & 23.16 & 42.80 & 9.90 & 27.36 & 10.94 & 28.16 \\
            & Llama-3.1-8B-Instruct & 33.38 & 42.47 & 5.86 & 9.08 & \underline{5.39} & 33.52 & 58.34 & 73.66 & 15.80 & 36.60 & 23.75 & 39.07 \\
            & DeepSeek-R1-Distill-Qwen-7B & 8.31 & 15.47 & 5.54 & 9.30 & 0.81 & 21.14 & 14.24 & 35.55 & 17.07 & 34.10 & 9.19 & 23.11 \\
        \midrule
        \multicolumn{14}{c}{\textbf{Listwise}$^{\tau}$} \\
        \midrule
        \multirow{7}{*}{\textbf{Listwise-Vanilla}}
            & Qwen3-8B & 29.88 & 37.77 & \underline{7.82} & 11.52 & 8.25 & 34.13 & 59.69 & 74.79 & 32.60 & 49.40 & 27.65 & 41.52 \\
            & gpt-oss-20b & 23.98 & 31.95 & \underline{7.82} & 10.98 & 4.16 & 26.55 & \underline{49.75} & \underline{67.53} & \underline{28.07} & \underline{56.00} & 22.76 & 38.60 \\
            & gemma-3-12b-it & 25.42 & 31.08 & \underline{7.82} & \underline{12.20} & 3.17 & 16.12 & 60.33 & 75.65 & 27.03 & 45.11 & 24.75 & 36.03 \\
            & Qwen3-8B-thinking & 31.57 & 38.78 & \underline{8.14} & 12.22 & 7.50 & 33.01 & \underline{59.92} & \underline{75.34} & 30.87 & 47.47 & 27.60 & 41.36 \\
            & Qwen2.5-7B-Instruct & 9.52 & 21.29 & 5.54 & 9.37 & 4.33 & 29.41 & 23.23 & 42.51 & \underline{10.10} & \underline{27.92} & 10.54 & 26.10 \\
            & Llama-3.1-8B-Instruct & 23.01 & 30.25 & 6.51 & 9.76 & 5.20 & 32.11 & 58.07 & 73.56 & 17.03 & 38.80 & 21.96 & 36.90 \\
            & DeepSeek-R1-Distill-Qwen-7B & 4.46 & 8.92 & 6.19 & 9.80 & 0.83 & 19.00 & 13.71 & 35.60 & 19.07 & 37.10 & 8.85 & 22.08 \\
        \midrule
        \multirow{7}{*}{\textbf{RankZephyr}}
            & Qwen3-8B & 45.06 & 53.74 & 7.49 & 12.09 & 9.09 & 35.75 & \textbf{61.35} & \textbf{76.38} & 31.60 & 48.83 & 30.92 & 45.36 \\
            & gpt-oss-20b & 32.17 & 43.00 & 6.84 & 9.07 & 1.99 & 17.19 & 41.31 & 60.11 & 21.23 & 46.03 & 20.71 & 35.08 \\
            & gemma-3-12b-it & 41.57 & 48.90 & 7.17 & 11.23 & 3.02 & 15.89 & \textbf{61.65} & \textbf{76.95} & 25.83 & 44.45 & 27.85 & 39.48 \\
            & Qwen3-8B-thinking & 43.61 & 51.33 & \underline{8.14} & 11.61 & 8.14 & 33.25 & 59.27 & 74.64 & 28.03 & 45.88 & 29.44 & 43.34 \\
            & Qwen2.5-7B-Instruct & 16.99 & 31.87 & \underline{5.86} & 9.40 & 4.10 & 30.57 & 23.35 & 43.12 & 8.83 & 26.93 & 11.83 & 28.38 \\
            & Llama-3.1-8B-Instruct & 34.58 & 43.71 & 6.51 & 9.30 & 5.06 & 33.26 & 58.00 & 73.98 & 14.53 & 36.74 & 23.74 & 39.40 \\
            & DeepSeek-R1-Distill-Qwen-7B & 7.71 & 14.45 & 6.19 & 9.24 & 0.58 & 18.48 & 13.80 & 36.46 & 17.43 & 35.31 & 9.14 & 22.79 \\
        \midrule
        \multicolumn{14}{c}{\textbf{Set Selection}} \\
        \midrule
        \multirow{7}{*}{\textbf{SetSelection-Vanilla}}
            & Qwen3-8B & 45.54 & 53.87 & \textbf{8.14} & \underline{12.76} & 8.89 & 35.05 & \underline{61.03} & \underline{75.76} & \underline{35.77} & \underline{51.58} & \underline{31.87} & \underline{45.80} \\
            & gpt-oss-20b & \textbf{42.41} & \textbf{51.61} & \underline{7.82} & \underline{11.87} & \textbf{9.14} & \textbf{37.57} & \textbf{54.37} & \textbf{72.64} & \textbf{39.10} & \textbf{62.83} & \textbf{30.57} & \textbf{47.30} \\
            & gemma-3-12b-it & \underline{46.63} & \underline{53.52} & 7.49 & 12.09 & 4.52 & 18.28 & \underline{61.47} & \underline{76.82} & \underline{29.40} & \underline{46.97} & \underline{29.90} & \underline{41.54} \\
            & Qwen3-8B-thinking & \underline{48.80} & 55.87 & \underline{8.14} & 12.52 & \underline{8.76} & \textbf{34.71} & \textbf{61.21} & \textbf{76.13} & \textbf{34.40} & \underline{50.19} & \textbf{32.26} & \underline{45.88} \\
            & Qwen2.5-7B-Instruct & \textbf{20.48} & \underline{34.91} & \textbf{6.84} & \underline{10.83} & 4.50 & 28.82 & \underline{29.43} & \underline{48.13} & \underline{10.10} & 27.47 & \underline{14.27} & \underline{30.03} \\
            & Llama-3.1-8B-Instruct & \underline{39.76} & \underline{48.77} & \underline{6.84} & \underline{10.46} & 4.68 & \underline{35.68} & \underline{59.96} & \underline{75.34} & \underline{33.10} & \underline{56.69} & \underline{28.87} & \underline{45.39} \\
            & DeepSeek-R1-Distill-Qwen-7B & \underline{15.90} & \underline{25.06} & \underline{6.84} & \underline{11.18} & \textbf{8.14} & \underline{34.62} & \textbf{36.04} & \underline{55.23} & \textbf{33.27} & \underline{50.31} & \underline{20.04} & \underline{35.28} \\
        \midrule
        \multirow{7}{*}{\textbf{SETR}}
            & Qwen3-8B & 29.64 & 33.71 & 5.86 & 9.10 & \textbf{9.59} & \textbf{36.89} & 53.95 & 67.42 & 32.83 & 49.58 & 26.37 & 39.34 \\
            & gpt-oss-20b & 20.60 & 26.38 & 5.21 & 7.91 & 3.11 & 22.03 & 38.03 & 54.57 & 21.73 & 46.18 & 17.74 & 31.41 \\
            & gemma-3-12b-it & 28.67 & 32.41 & 5.86 & 9.55 & \underline{4.62} & \underline{19.55} & 54.14 & 68.36 & 26.67 & 44.97 & 23.99 & 34.97 \\
            & Qwen3-8B-thinking & 28.92 & 33.15 & 5.86 & 9.17 & 8.41 & 34.51 & 52.75 & 66.29 & 29.80 & 47.22 & 25.15 & 38.07 \\
            & Qwen2.5-7B-Instruct & 13.98 & 23.17 & 4.56 & 8.29 & \textbf{4.77} & \underline{30.80} & 25.68 & 42.85 & \underline{10.10} & 27.68 & 11.82 & 26.56 \\
            & Llama-3.1-8B-Instruct & 25.06 & 30.33 & 5.54 & 8.64 & 5.08 & 35.43 & 53.49 & 67.51 & 23.47 & 46.16 & 22.53 & 37.61 \\
            & DeepSeek-R1-Distill-Qwen-7B & 11.57 & 17.83 & 4.56 & 8.23 & 6.19 & 31.70 & 31.16 & 48.83 & 27.07 & 43.65 & 16.11 & 30.05 \\
        \midrule
        \multirow{7}{*}{\textbf{Rank4Gen}}
            & Qwen3-8B & \textbf{46.51} & \textbf{55.91} & \textbf{8.14} & \textbf{14.12} & 9.28 & 36.23 & 59.80 & 75.26 & \textbf{35.83} & \textbf{52.84} & \textbf{31.91} & \textbf{46.87} \\
            & gpt-oss-20b & \underline{38.31} & \underline{50.42} & \textbf{8.47} & \textbf{12.19} & \underline{5.05} & \underline{28.69} & 48.96 & 66.79 & 27.27 & 53.44 & \underline{25.61} & \underline{42.31} \\
            & gemma-3-12b-it & \textbf{46.87} & \textbf{54.88} & \textbf{8.47} & \textbf{14.01} & \textbf{5.43} & \textbf{20.64} & 61.35 & 76.31 & \textbf{31.70} & \textbf{50.64} & \textbf{30.76} & \textbf{43.30} \\
            & Qwen3-8B-thinking & 47.71 & \textbf{57.66} & \textbf{9.12} & \textbf{14.21} & \textbf{8.83} & \underline{34.54} & 57.77 & 73.49 & \underline{32.93} & \textbf{50.61} & \underline{31.27} & \textbf{46.10} \\
            & Qwen2.5-7B-Instruct & \underline{19.40} & \textbf{36.18} & \textbf{6.84} & \textbf{11.76} & \underline{4.52} & 28.83 & \textbf{32.85} & \textbf{50.78} & \textbf{11.23} & \textbf{28.80} & \textbf{14.97} & \textbf{31.27} \\
            & Llama-3.1-8B-Instruct & \textbf{42.05} & \textbf{52.86} & \textbf{7.49} & \textbf{12.05} & \textbf{5.74} & \textbf{37.66} & \textbf{60.10} & \textbf{76.09} & \textbf{33.47} & \textbf{58.21} & \textbf{29.77} & \textbf{47.37} \\
            & DeepSeek-R1-Distill-Qwen-7B & \textbf{18.43} & \textbf{31.14} & \textbf{8.14} & \textbf{13.06} & \underline{7.29} & \textbf{35.62} & \underline{34.91} & \textbf{55.80} & \underline{33.17} & \textbf{51.16} & \textbf{20.39} & \textbf{37.36} \\
        \bottomrule
    \end{tabular}}
   \caption{Main results on five RAG benchmarks using different rankers and generators. Performance is reported in Exact Match (EM) and token-level F1. $\tau$ denotes that the top-10 retrieved documents are used. The best and second-best results are highlighted in \textbf{bold} and \underline{underline} for each generator, respectively.}
    \label{tab:full_results}
\end{table*}